\documentclass{jfm}

\usepackage{graphicx}
\usepackage{epstopdf}
\usepackage{natbib}
\usepackage{bm}
\usepackage{amsmath}
\usepackage{amsfonts}
\usepackage{amssymb}
\usepackage{natbib}
\usepackage[utf8]{inputenc}
\usepackage{color} 
\usepackage[dvipsnames]{xcolor}
\usepackage{color,soul}

\ifCUPmtlplainloaded \else
  \checkfont{eurm10}
  \iffontfound
    \IfFileExists{upmath.sty}
      {\typeout{^^JFound AMS Euler Roman fonts on the system,
                   using the 'upmath' package.^^J}%
       \usepackage{upmath}}
      {\typeout{^^JFound AMS Euler Roman fonts on the system, but you
                   dont seem to have the}%
       \typeout{'upmath' package installed. JFM.cls can take advantage
                 of these fonts,^^Jif you use 'upmath' package.^^J}%
      }
  \else
  \fi
\fi

\ifCUPmtlplainloaded \else
  \checkfont{msam10}
  \iffontfound
    \IfFileExists{amssymb.sty}
      {\typeout{^^JFound AMS Symbol fonts on the system, using the
                'amssymb' package.^^J}%
       \usepackage{amssymb}%

      }{}
  \fi
\fi

\ifCUPmtlplainloaded \else
  \IfFileExists{amsbsy.sty}
    {\typeout{^^JFound the 'amsbsy' package on the system, using it.^^J}%
     \usepackage{amsbsy}}
    {}
\fi

\newsavebox{\astrutbox}
\sbox{\astrutbox}{\rule[-5pt]{0pt}{20pt}}

\title[Dependance of acoustic streaming in droplets on viscosity]{Influence of viscosity on acoustic streaming in sessile droplets: an experimental and a numerical study with a Streaming Source Spatial Filtering (SSSF) method.}

\author[A. Riaud, M. Baudoin, O. Bou Matar, J.-L. Thomas and P. Brunet]
{A. Riaud $^{1,2}$, \ns M. Baudoin $^{1,}$\thanks{Email address for correspondence: michael.baudoin@univ-lille1.fr / Web site: http://films-lab.univ-lille1.fr/michael/},\ns O. Bou Matar $^1$, \break J.-L. Thomas $^2$, \ns and P. Brunet  $^{3}$}

\affiliation{$^1$Institut d'Electronique, de Micro\'electronique et Nanotechnologie (IEMN), LIA LEMAC/LICS, Universit\'{e} Lille 1, UMR CNRS 8520, Avenue Poincar\'e, 59652 Villeneuve d'Ascq, France \\[\affilskip]
$^2$Institut des NanoSciences de Paris (INSP), Universit\'{e} Pierre et Marie Curie, UMR CNRS 7588, 4 place Jussieu, 75252 Paris cedex 05, France\\[\affilskip]
$^3$Laboratoire Mati\`ere et Syst\`emes Complexes UMR CNRS 7057, 10 rue Alice Domon et L\'eonie Duquet 75205 Paris Cedex 13, France }

\pubyear{2010}
\volume{650}
\pagerange{119--126}
\date{XX revised XX accepted XX - To be entered by editorial office}
\begin{document}

\maketitle

\begin{abstract}
When an acoustic wave travels in a lossy medium such as a liquid, it progressively transfers its pseudo-momentum to the fluid, which results in a steady acoustic streaming. Remarkably, the phenomenon involves a balance between sound attenuation and shear, such that viscosity vanishes in the final expression of the velocity field. For this reason, the effect of viscosity has long been ignored in acoustic streaming experiments. Here, we show experimentally that the viscosity plays a major role in cavities such as the streaming induced by surface acoustic waves in sessile droplets. We develop a numerical model based on the spatial filtering of the streaming source term
to compute the induced flow motion with dramatically reduced computational requirements. We evidence that acoustic fields in droplets are a superposition of a chaotic field and a few powerful caustics. It appears that the caustics drive the flow, which allows a qualitative prediction of the flow structure. Finally, we reduce the  problem to two dimensionless numbers related to the surface and bulk waves attenuation and simulate hemispherical sessile droplets resting on a lithium niobate substrate for a range of parameters. Even in such a baseline configuration, we observe at least four distinct flow regimes. For each of them, we establish a correlation of the average streaming speed in the droplet, which is increasingly dependent on the bulk wave attenuation as the viscosity increases. These correlations extend our results to a wide range of fluids and actuation frequencies.
\end{abstract}

\begin{keywords}
Authors should not enter keywords on the manuscript, as these must be chosen by the author during the online submission process and will then be added during the typesetting process (see http://journals.cambridge.org/data/\linebreak[3]relatedlink/jfm-\linebreak[3]keywords.pdf for the full list)
\end{keywords}

\section{Introduction \label{intro}}

Two incommensurable time scales are involved when sound waves propagate in a fluid: the frequency of the sound and the characteristic time of the flow evolution. This distinction defines the boundary between acoustics and hydrodynamics. It happens in practice that some physical phenomena overlap this frontier and challenge researchers from both fields, like shock waves and supersonic hydrodynamics, but also noisy powerful hydrodynamic turbulence and steady flows induced by intense sound beams: namely the acoustic streaming. 

Thanks to \cite{ptrsl_rayleigh_1884} and \cite{Eckart} pioneering works, it is now well established that acoustic streaming reveals the momentum transfer from the wave to the fluid by sound attenuation. Most authors (\cite{Lighthill1978391,Mitome1998,Westerwelt1953,arfm_ryley_2001,Riley_TCFD,ap_nyborg_1998,loc_wiklund_2012}) distinguish two types of streaming depending if the damping arises from shear stress on the walls (boundary layer-driven streaming) or from viscous dissipation in the bulk (Eckart streaming). \cite{Rednikov2011} have proven that the former acts as an efficient slip velocity outside a thin Stokes boundary layer, while \cite{Lighthill1978391} has provided a convenient body force expression to account for the latter. In what follows, we will treat exclusively the case of Eckart (bulk) streaming, which is relevant for geometries much larger than the acoustic wavelength (\cite{Vanneste2011}). Quickly following Eckart's theoretical work (\cite{Eckart}), \cite{Liebermann1949} experimentally proved that the attenuation of freely-propagating sound waves was mostly due to the bulk viscosity, a frequency-dependent parameter combining hard sphere collision integral and chemical reactions kinetics. As emphasized by \cite{Eckart}, the hydrodynamic forcing term is proportional to the sound attenuation, which itself varies linearly with the viscosity.

Here appears one of the greatest paradox of acoustic streaming: although the momentum source for the fluid is proportional to the viscosity, it mostly dissipates this momentum through shear stress, such that streaming velocity is expected to be independent of viscosity. Experimentally, it has been confirmed that acoustic streaming occurs for a wide range of fluids from superfluid Helium (\cite{Rooney1982}) to very viscous polymers (\cite{Mitome1998}). Nevertheless, this assertion must be mitigated for two reasons: (i) at large sound intensity or low viscosity (\cite{Lighthill1978391}), hydrodynamic momentum convection becomes the main dissipation mechanism resulting in a velocity slope break marking the transition between slow and fast acoustic streaming (\cite{Liebermann1949,Kamakura1995}) (ii) at high viscosity or high frequency, the sound wave attenuates quickly confining the forcing term to a smaller region of space (\cite{ap_nyborg_1998}), which has recently been experimentally evidenced (\cite{pre_dentry_2014}). Although the first effect has been studied experimentally and numerically by \cite{Kamakura1995} and \cite{Matsuda1996763}, the second one has received little attention and leads to many misunderstandings. 

\begin{figure} \begin{center}
\includegraphics[width=0.5\textwidth]{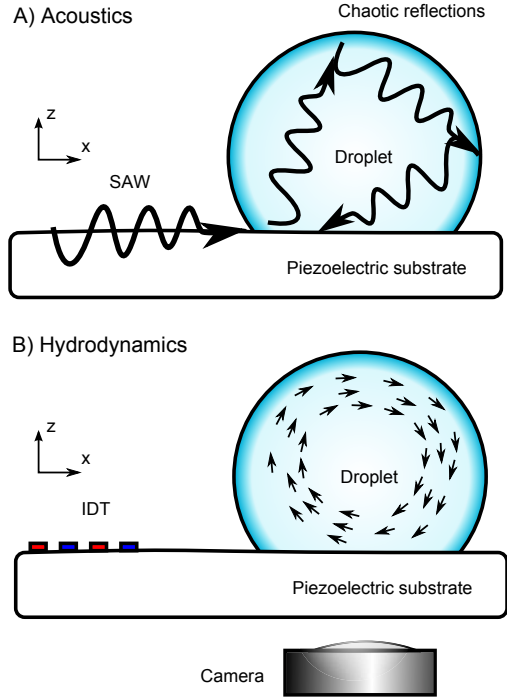}
\caption{A sessile droplet rests on a piezoelectric substrate. \textbf{A) Acoustics.} A SAW propagating at the surface of the solid radiates in the liquid. It is reflected a great number of times at the liquid-solid and liquid-air interfaces, resulting in a complex standing wave pattern. \textbf{B) Hydrodynamics} As the wave propagates in the liquid, it dissipates some momentum which surprisingly generates a steady flow with large-scale eddies.}
\label{fig: physical_principal}
\end{center} \end{figure}

High frequency sound waves and large viscosity liquids are routinely used in microfluidics (see e.g. \cite{rmp_friend_2011,loc_wiklund_2012}). Indeed, contactless robust fluid actuation for a wide range of liquids is a primary requirement for this emergent discipline, and miniaturized acoustical sources such as interdigitated transducers are readily available. Herein, a problem of considerable interest is the acoustic streaming induced by surface acoustic waves (SAW) in sessile droplets, as illustrated on figure \ref{fig: physical_principal}. An oscillating voltage applied on an interdigitated transducer generates a SAW at the surface of a piezoelectric medium. This wave propagates almost unattenuated until it meets a liquid droplet. As it moves below the liquid, the surface oscillations are damped by the inertial stress of the fluid, and the surface wave gradually leaks in the liquid, generating bulk acoustic waves. 

For increasing SAW power, one can achieve droplet mixing (\cite{apl_sritharan_2006,prl_frommelt_2008} or centrifugation \cite{Bourquin2010}), displacement (\cite{abc_wixforth_2004,saa_renaudin_2006,pre_brunet_2010,Alzuaga2005,apl_baudoin_2012,Tomohiko2015}), division (\cite{Zhang_Anliang2013,Collignon2015,ieeetuffc_riaud_2016}), heating (\cite{ieeeus_kondoh_2005,Beyssen2006,jjap_ito_2007,saa_kondoh_2009,Roux-Marchand2012,pnas_reboud_2012,ieeetuffc_rouxmarchand_2015,afm_shilton_2015}) and finally jetting (\cite{jjap_shiokawa_1990,prl_tan_2009}) or atomization (\cite{pof_yeo_2008}) depending on the droplet size (\cite{prl_tan_2009}). These phenomena are still only partially understood and the underlying physics is sometimes subject to some controversy. For instance, most authors agree that the wave momentum is transfered to the fluid, but some argue that the momentum transfer happens in the bulk by acoustic streaming (\cite{prl_tan_2009,Alghane2012,Schindler2006}), while others point out that sound reflections on a fluid interface also generates a measurable surface stress (\cite{Hertz_Mende}) called acoustic radiation pressure (\cite{Mitome1998,Sato_and_Fujii,stanzial2003}) which could contribute to the aforementioned effects (\cite{Alzuaga2005,pre_brunet_2010}). Even though, there is a general consensus on the droplet mixing which can happen without significant free-surface deformation and is therefore widely attributed to acoustic streaming.

Acoustic streaming in sessile droplets represents a significant overlap between acoustics and hydrodynamics, and researchers from both fields used their own approach to develop a better understanding of the phenomenon. In the hydrodynamic viewpoint, acoustic streaming is of considerable interest since it allows contactless generation of vorticity and fluid mixing, especially in microfluidics (\cite{rmp_friend_2011,loc_wiklund_2012}). Conversely, the acoustic community dedicated little work to this specific phenomenon. Consequently, although the propagation of surface acoustic waves is extremely well understood (see e.g. \cite{Royer1,Royer2}), the intermediate step between the leaky SAW radiation and the hydrodynamic flow remains unclear, and the droplet appears essentially as an acoustical blackbox. 

Hydrodynamic studies on droplet acoustic streaming at megahertz frequencies started in 1990 with \cite{jjap_shiokawa_1990} seminal paper. They performed several experiments of droplet displacement, jetting and atomization using surface acoustic waves at 50 MHz. In the same paper, Shiokawa and coworkers lay down several important theoretical foundations for subsequent studies. The authors observed the formation of jets when exposing water droplets to high power SAWs at 50 MHz. In their experiments, the liquid is ejected in the same direction as if the droplet was an unbounded medium. According to this observation, Shiokawa et al. assumed that the acoustic field in water droplets could be reduced to the incident field and proposed to neglect the internal reflections of the wave. This postulate allowed him to use Nyborg's expression of the acoustic streaming force in order to compute the order of magnitude of the acoustic streaming in sessile droplets. Finally, Shiokawa and his coauthors emphasized that the gigantic attenuation of the leaky SAW beneath the droplet exceeds by far the viscous attenuation of the same sound wave in the droplet bulk, see also (\cite{Cheeke}). Thus, their calculations were performed in the inviscid approximation for sound waves. Most subsequent works followed the guidelines of Shiokawa, neglecting the internal reflections of the acoustic wave on the droplet surface and using Nyborg's force expression. 

Experimental and theoretical studies in the continuity of Shiokawa include  \cite{Du2009} who observed droplet acoustic streaming at 62.4 and 128 MHz and \cite{Alghane2011,Alghane2012} work (experiments performed at 60 MHz). In these studies, the magnitude of the SAW was unknown and set by a least square curve fitting. Droplet deformation due to high power surface acoustic waves was numerically investigated by \cite{Schindler2006} but they neglected the details of the SAW radiation in the drop and had to make assumptions similar to Shiokawa's. In another work, \cite{Raghavan2010} observed the flow induced by surface acoustic waves in sessile droplets at frequencies of 20 MHz. Their study departs significantly from Shiokawa guidelines by including a two-dimensional ab-initio numerical simulation where they solve the stationary compressible Navier-Stokes equation, including the acoustic field. Contrasting with earlier studies, the magnitude of the SAW displacement was known, leaving no room for adjusting parameters. Although he recovered the correct flow pattern, Raghavan reported fluid velocities an order of magnitude below what was measured experimentally. \cite{koster2007} proposed an algorithm to compute the flow in a sessile droplet exposed to surface acoustic wave. The method includes the sound propagation and the droplet deformation, and works iteratively (refreshing the acoustics after a certain number of hydrodynamic time steps). Despite the outstanding nature of the study, the emphasis is put on the method rather than the physical results. K\"oster's investigation does not provide any clear comparison to experimental results and the numerical study is limited to one special size of droplet excited at one specific frequency. 

In 2011, \cite{Vanneste2011} pointed out that most numerical studies based on Nyborg's expression as in Shiokawa's work relied on inviscid formulations of the sound wave equation, which could not generate vorticity. Since incompressible flows are inherently vortical flows, earlier numerical studies were put at stake by Vanneste and Buhler's assertion. In order to remedy to the situation, they developed a rigorous analytical computation of the streaming generated by surface acoustic waves in a square cavity based on vorticity conservation. In their analysis, the box was transparent to acoustic waves, which is similar to Shiokawa \textit{et al}'s analysis of neglecting internal reflections. Another important contribution of Vanneste was to single out the bulk streaming (Eckart streaming) against boundary layer streaming as the flow motor in the case of cavities much larger than the wavelength, which is the case for millimetric droplets irradiated by SAW of frequency larger than 10 MHz.

The acoustic foundations of Shiokawa's framework were also challenged by \cite{pre_brunet_2010} who tested the validity of the reflection-free droplet assumption. In this work, the authors simulated the acoustic field in a two dimensional droplet. They found out that the field in water droplets exposed to 20 MHz SAW was extremely complicated and showed little coherent structure. Nevertheless, for attenuations about 100 times larger than water, the incident wave accounted for most of the acoustic field. Another study performed by \cite{Quintero2013} at 3.5 MHz revealed the acoustic field in three dimensions in the low frequency range where the wavelength is comparable to the droplet size. Again, no clear structure was present. In these two studies, the knowledge of the acoustic field was not used to proceed to the next step and compute the streaming forcing term.

At present, our understanding of the acoustic streaming in sessile droplets faces the three following issues: What is the acoustic field in the droplet ? Since we infer the field to be quite complicated, how does it generates some coherent flow pattern ? How to compute this flow while ensuring vorticity conservation ? In the continuity of \cite{pre_brunet_2010}, we tune the liquid viscosity to explore the gap between Shiokawa hypothesis of reflection-free droplets and actual droplet experiments. In section \ref{sec: Experiment}, we present an experimental study of acoustic streaming in droplets of different viscosities, and show a transition of flow pattern from four to two eddies for increasing viscosity. This contradiction with the inviscid viscosity appeals for an in-depth review of the acoustic streaming theoretical foundations, exposed in section \ref{sec: Theory}. We single out the dominant inviscid term that does not contribute to vorticity creation and extract Lighthill's acoustic streaming driving force (\cite{Lighthill1978391}). We use this expression in section \ref{sec: Numerical model}, where we detail a numerical algorithm to compute the acoustic field in the droplet, deduce the streaming forcing term and then reproduce the 3D flow pattern observed experimentally based on Large Eddy Simulation. Section \ref{sec: DimensionlessAnalysis} opens a discussion by comparing numerical and experimental results. We show that simple arguments of geometrical acoustics and sound attenuation can provide a qualitative prediction of the flow topology. Finally, we reduce the droplet outer streaming to a two-parameters non-dimensional problem and provide a correlation to extend our results to many fluids and actuation frequency. Provided the wavelength is much shorter than the droplet size, our approach (not restricted to plane waves) allows simulating droplet streaming at low SAW actuation frequency and then extrapolate the results to higher frequencies. This considerably alleviates memory requirements to simulate acoustic streaming in complex geometries.
 



\section{Experiments} \label{sec: Experiment}

\subsection{Experimental Setup}
Surface acoustic waves were generated at the surface of a X-cut lithium niobate piezoelectric substrate in the Z-direction by interdigitated electrodes, with a spatial period of 175 $\mu$m corresponding to a resonant frequency of 19.9 MHz (the sound speed in this direction is $3484$ ms$^{-1}$, see \cite{CampbellJones}). In practice, the best actuation efficiency was obtained at 20.37 MHz which was used as the driving frequency for all experiments.  A  water-glycerol droplet of $12.5$ $\mu$L was placed on the substrate initially treated with OTS Self Assembled Monolayer (SAM) to obtain hydrophobic wetting properties (see figure \ref{fig: physical_principal}). The water-glycerol mixture was used to tune the shear and bulk viscosities with relatively weak variations of the other relevant driving parameters (physical data are shown in table \ref{tab: physical parameters}). Spherical latex beads of 10 $\mu$m diameter coated with fluorescent dyes (ThermoScientific) were dispersed in the droplet prior to experiments to visualize its inner-flow. We minimized the droplet evaporation by deporting the light source with an optical fiber, using the cold part of the optical spectrum and restraining the experiment duration below 2 minutes. The images were acquired via an Hamamatsu high resolution camera and quantitative velocity magnitude was measured using the PIV module of ImageJ (https://sites.google.com/site/qingzongtseng/piv). We restricted the power of the SAW to a few tens of picometers to minimize the droplet deformation, which are observed at much higher amplitude (\cite{pre_brunet_2010,Schindler2006,Alghane2012,apl_baudoin_2012}). The substrate vertical amplitude of vibration on the central finger of the IDT was calibrated for the range of actuation power used in the present experiments with a laser Doppler vibrometer (SH130, B.M. Industries).

\begin{table}
\begin{center}
\begin{tabular}{ccccccccc}
$w_\mathtt{glyc.}$ & $x_\mathtt{glyc.}$  & $\mu$ (mPa.s) & $c_0$ (m/s) & $\rho_0$ (kg/m$^3$)  & $b = 4/3 + \xi/\mu$  & $\theta_c$ $(^o)$ & $\Lambda = \frac{D\omega^2\nu b}{c^3}$ & $Re_{ac} =  \frac{\rho_0 c_0 ^2}{\omega \mu b}$\\ 
0.00 &  0.00 & 0.892 &  1510 & 1000 & 4.53 & 88 & 0.068 & 4490 \\ 
0.10 & 0.02 & 1.15 &  1540 & 1020 &  4.45 & 87 & 0.080 &  3716 \\ 
0.20 &  0.05 & 1.52 & 1580 & 1050 & 4.34 & 86 & 0.095 & 3162  \\
0.30 & 0.08 &  2.12 & 1630 & 1070 & 4.22 & 85 & 0.11 &  2528\\ 
0.40 & 0.11 &  3.13 & 1680 & 1100 & 4.04 & 84 & 0.15 & 1953 \\ 
0.50 &  0.16 & 5.00 & 1720 &  1130 & 3.75 & 83 & 0.20 &  1418 \\ 
0.60 & 0.23 & 8.85 & 1780 & 1150 & 3.49 & 82 & 0.29 &  938 \\ 
0.70 & 0.31 & 18.1 & 1830 & 1180 & 3.34 & 81 & 0.52 & 520  \\
0.80 & 0.44 & 45.4 & 1880 & 1210 & 3.03 & 80 & 1.0 &  247 \\ 
0.90 & 0.64 & 156 &  1910 & 1230 & 2.50 & 78 & 2.80 & 91 \\  
\end{tabular} 
\\ $c_s = 3484$ m/s, $\rho_s = 4650$ kg/m$^3$, $1.7 < \alpha D = 3.7\times 10^{-9} \omega \rho_0 D < 2.3$ \\
\end{center}
\caption{Physical properties of lithium niobate and water-glycerol mixtures at 25$^o$C for different mass fraction $w_\mathtt{glyc}$ and thus volume fraction $x_\mathtt{glyc}$ of glycerol\label{tab: physical parameters}. Data for the viscosity $\mu$ of the water-glycerol mixture are extracted from \cite{Cheng2008} paper, while the sound speed $c_0$, the density $\rho_0$, the bulk viscosity $\xi$ (and thus the coefficient $b$) are  extracted from \cite{Slie1966} paper. The sound speed $c_s$ of Rayleigh waves in $X$-cut niobate lithium in the Z-direction is extracted from \cite{CampbellJones} and the density $\rho_s$ is available in handbooks. Finally $\Lambda$ is a dimensionless parameter characterizing the transmission efficiency of the Rayleigh wave to the liquid. Droplet base diameter can be computed as follows: $D_{base} = 2 \sin(\theta_c) \left[\frac{3 V}{\pi(2 - 3 \cos(\theta_c) + \cos^3(\theta_c))}\right]^{1/3}$. Accordingly, all droplets diameter in our experiments range between 3.7 and 4.0 mm.}
\label{table1}
\end{table}

\begin{figure*} \begin{center}
\includegraphics[width=\textwidth]{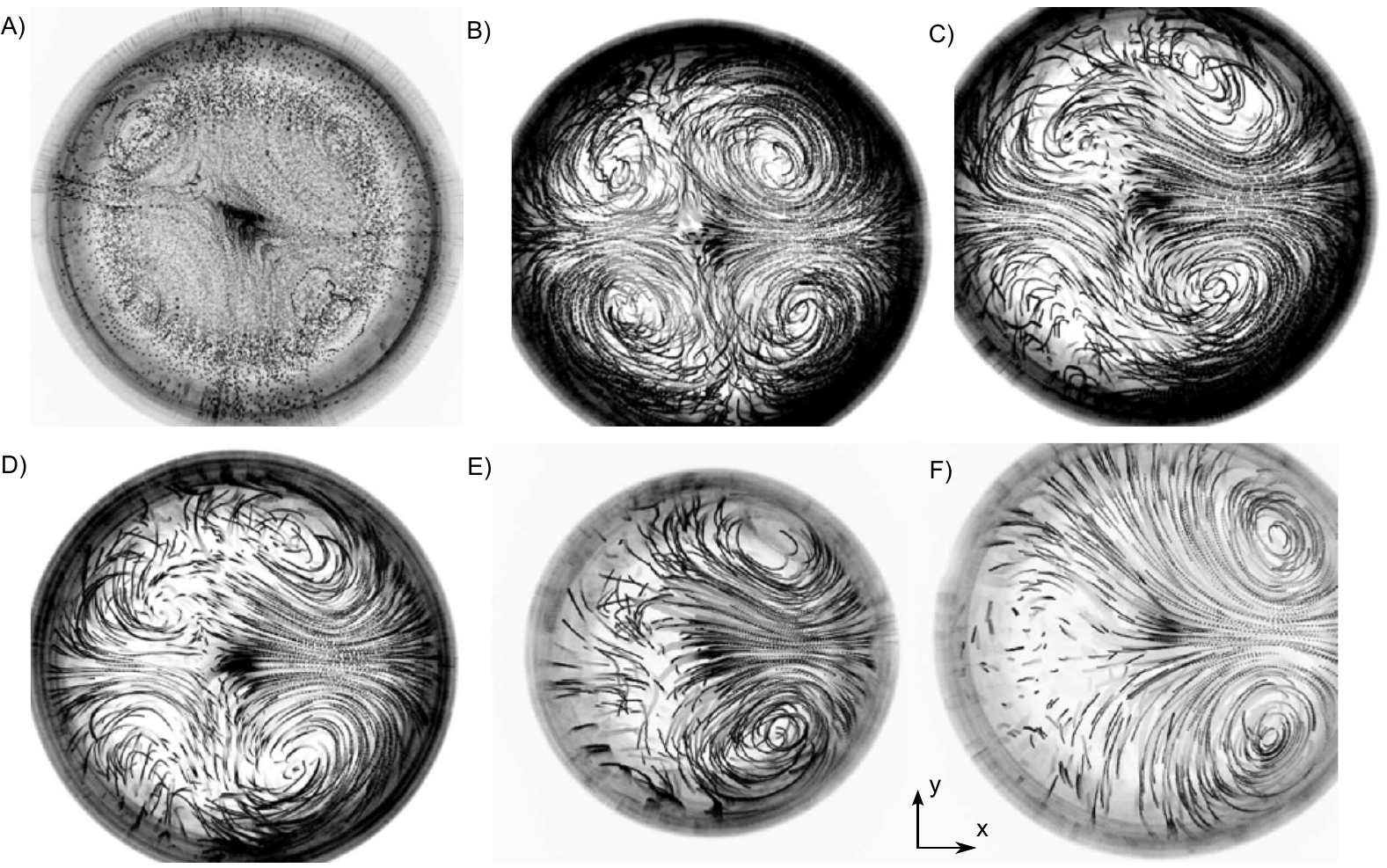}
\caption{Flow visualization from below, at various glycerol concentration. The SAW propagates from left to right with an amplitude $u_0\simeq 62$ pm. $V_{droplet}$ = 12.5 $\mu$l and the magnitude of the acoustic perturbation displacement $u_0\simeq 62$ pm. As the viscosity increases, one remarks the progressive transition from a four-vortex to a two-vortex flow structure. (A) Pure water (B) 30 wt$\%$ glyc. (C) 40 wt$\%$ glyc. (D) 60 wt$\%$ glyc. (E) 80 wt$\%$ glyc. (F) 90 wt$\%$ glyc.}
\label{fig: experimental_top_view}
\end{center} \end{figure*}

Finally, the droplets inner flow was visualized from below (to avoid optical aberrations by the drop surface) with a Hamamatsu high resolution camera mounted on an inverted microscope (Olympus IX71). The depth of field is estimated to be 16 $\mu$m for objects of the size of a pixel when using the 4$\times$ magnification objective with this inverted microscope, allowing the vizualisation of a droplet cross-section. The vertical position of the cut was adjusted by eye as close as possible to half of the drop height, although this condition is achieved within a few percent accuracy. After turning on the SAW generator, we waited for the droplet flow pattern to reach a steady state. This duration varied widely with viscosity, from seconds for water droplets up to minute for the most viscous mixtures (as expected from theoretical analysis). Figure \ref{fig: experimental_top_view} reproduces a few examples of such particles trajectories viewed from below, obtained at increasing viscosities from (A) to (F). The streamlines are obtained by a simple superposition of successive images.

The velocity field in the representative cut was extracted from the pictures of the flow streams presented above by using particle image velocimetry (PIV) ImageJ plugin (see figure \ref{fig: PIV bottom view}). The analysis was further refined by discarding the 5\% least reliable velocity vectors \footnote{The reliability criteria was the magnitude of the Laplacian of the velocity field.}. We made sure that the system had reached steady state by waiting until the space-averaged magnitude of the velocity field did not vary by more than 10\% between two different time intervals.  \textit{In fine}, each couple of images provides a flow map, and each plot in figure \ref{fig: PIV bottom view} is the average of three different flow maps obtained with the same droplet at steady state. Then the average velocity was determined by averaging spatially the flow field in the drop. The streaming pattern being inhomogeneous in space, this space-averaged flow velocity over one cross-section only gives an order of magnitude of the volume-averaged velocity. The resulting trend is presented in the discussion section together with numerical results (see fig. \ref{fig: quantitative flow speed}).

\subsection{Results}

These experimental results (see figure \ref{fig: experimental_top_view}) show unambiguously that the streaming flow pattern in a droplet excited by SAWs depends on the fluid viscosity. Fluids of increasing viscosity lead to progressive loss of left/right symmetry (hence along the direction of propagation of the SAW). The situation at low viscosity (up to 30 wt$\%$ glycerol - figure \ref{fig: experimental_top_view}-(A) and (B)) shows 2 pairs of vortices, both at the rear and the front of the drop (with respect to the direction of wave propagation). As the viscosity increases beyond a few times that of water (figure \ref{fig: experimental_top_view}-(C) and (D)) the front vortices start shrinking, while at even higher viscosity (above 80 wt$\%$ glycerol, figure \ref{fig: experimental_top_view}-(E) and (F)) the front vortices have completely disappeared. Counterintuitively, complex eddies are observed at the opposite side of the excitation at the highest viscosity (corresponding to very low Reynolds numbers)  (figure \ref{fig: experimental_top_view} E-F).

\begin{figure} \begin{center}
\includegraphics{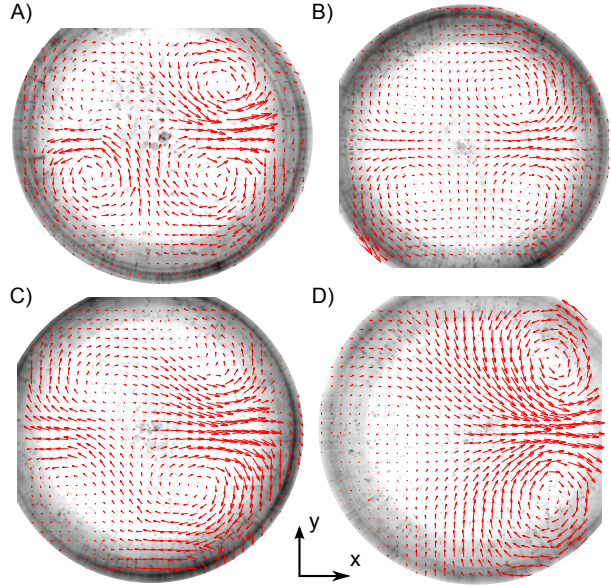}
\caption{Experimental velocity field from below, at various viscosities. The SAWs propagate from left to right with an amplitude $u_0\simeq 62$ pm. $V_{droplet}$ = 12.5 $\mu$l. As the viscosity increases, one remarks the progressive transition from a four-vortices to a two-vortices flow structure. (A) 10 wt$\%$ glycerol ($\mathcal{U}_{max} \simeq 180$ $\mu$m/s) (B) 30 wt$\%$ glycerol ($\mathcal{U}_{max} \simeq 100$ $\mu$m/s) (C) 40 wt$\%$ glycerol ($\mathcal{U}_{max} \simeq 70$ $\mu$m/s) (D) 90 wt$\%$ glycerol ($\mathcal{U}_{max} \simeq 10$ $\mu$m/s). The arrow length is indicative of the velocity magnitude for each experiment.}
\label{fig: PIV bottom view}
\end{center} \end{figure}

Then PIV measurements (figure \ref{fig: PIV bottom view}, magnitude are indicated in the caption) show that larger viscosities (from 1.15 mPa.s to 156 mPa.s)  are associated with a decreasing velocity magnitude (from 180 $\mu$m/s to 10 $\mu$m/s). This is in contradiction with the widespread assumption of a viscosity-independent streaming velocity. 

In order to unveil the underlying physics, we performed a systematic comparison between models, simulations and experiments of the flow pattern and average velocity in the drop for different viscosities. In the next sections, we therefore describe an adequate theory, introduce a numerical method allowing the computation of the 3D streaming flow in the drop with dramatically reduced numerical cost and perform a comparison with experiments to achieve a comprehensive understanding of the whole process behind the acoustic streaming in sessile droplets.

\section{Theory} \label{sec: Theory}

In this section, we re-establish acoustic streaming constitutive equations. At first we introduce a relevant field decomposition into periodic fluctuations (corresponding to the acoustic wave) and time averaged terms (corresponding to the acoustic streaming). Then, from the compressible Navier-Stokes equations, we derive a constitutive nonlinear equation for each of these contributions. In the latter appears a force under the form of a nonlinear combination of acoustic terms, which drives the acoustic streaming and the acoustic radiation pressure (\cite{Gusev1979,Mitome1998}). This driving force is recast in the last section as a convenient expression based on the sum of a conservative force plus a quantity proportional to the Poynting vector.

\subsection{Field decomposition}

As stated in the introduction, we can resolve each physical quantity $f$ into three contributions: hydrostatics $f_0$, acoustics $\tilde{f_1}$ and hydrodynamics $\bar{f_2}$. They represent respectively the system at rest (without acoustic field), the oscillating part of the perturbation induced by sound waves and the time averaged part of the perturbation over an acoustic period. In our experiments, acoustic and hydrodynamic Mach numbers are small. Moreover, solid displacements hardly exceed 0.5 nm, which restricts acoustic perturbation velocity magnitude below $10$ mm/s and consequently streaming velocities below $1$ mm/s. Accordingly, the fluid density $\rho$ the pressure $p$ and the Eulerian velocity $v$, can be expressed as follows: 
\begin{eqnarray}
\rho = \rho_0 + \tilde{\rho}_1+\bar{\rho}_2,\\
p = p_0 + \tilde{p}_1+\bar{p}_2,\\
v_i = \tilde{v}_{1,i}+\bar{v}_{2,i}\\
v_{0,i} = 0.
\end{eqnarray}
with $i\in\{x,y,z\}$, $\bar{f_2} = \langle f - f_0 \rangle$, $\langle \rangle$ the time average, $\tilde{f_1} = f - f_0 - \bar{f_2}$, $\langle \tilde{f_1} \rangle = 0$, and $\bar{f_2} \ll \tilde{f_1} \ll f_0$. The low Mach numbers assumption gives $\tilde{v}_1,\bar{v}_2 << c_0$, with $c_0$ the sound speed in the fluid at hand. To simplify the notations, the indices $1$ and $2$ will be omitted in the following.

\subsection{Fundamental Equations}
\label{sect: Fundamental Equations}

The starting point of the derivation are the compressible isentropic Navier-Stokes equations. These equations are relevant to compute acoustic streaming in liquids since, in this case, thermal effects (wave thermal damping, fluid heating) can be neglected compared to their viscous counterpart (viscous damping, acoustic streaming). Indeed, thermal effects are proportional to $(\gamma - 1)$, with $\gamma$ the adiabatic index and thus are very weak in liquids (see e.g. \cite{ja_coulouvrat_1992}). 

In this case, the mass conservation equation for a fluid reads:
\begin{equation}
\partial_t\rho + \partial_i\rho v_i = 0,
\label{eq:mass}
\end{equation}
and the momentum conservation equation:
\begin{equation}
\partial_t\rho v_i + \partial_j (\rho v_i v_j) = -\partial_i p + \mu\partial^2_{jj} v_i + \left(\frac{\mu}{3}+\xi \right)\partial^2_{ij}v_j.
\label{eq:momentum}
\end{equation}
In these equations, $\mu$ stands for the dynamic viscosity, $\xi$ for the bulk viscosity, $t$ for the time and the indices $i$ and $j$ follow Einstein summation convention.
The second-order isotropic Taylor-expansion of the equation of state reads:
\begin{equation}
dp = c_0^2 d\rho + \frac{1}{2}\Gamma  d\rho^2,
\label{eq:state}
\end{equation}
with $\Gamma = \left. \frac{\partial^2 p }{\partial \rho^2} \right\rvert_s = \frac{Bc_0^2}{A\rho_0}$. $A$ and $B$ are two nonlinear coefficients classically introduced in nonlinear acoustics.

\subsection{Time averaged equations at second order: acoustic steady streaming}
\label{sect: Time avergaed}

If we take the time average of the mass and momentum conservation equations (\ref{eq:mass}) and (\ref{eq:momentum}) up to second order, and introduce the Poynting vector (also called intensity vector in the field of acoustics) $\Pi_i = \tilde{p}\tilde{v}_i$, we get:
\begin{equation}
\partial_t \bar{\rho} + \rho_0 \partial_i \bar{v}_i + \frac{1}{c_0^2}\partial_i \left\langle \Pi_i \right\rangle = 0,
\label{eq:massa}
\end{equation}
and :
\begin{equation}
\partial_t \left( \rho_0 \bar{v_i} + 1/c_0^2 \left\langle \Pi_i \right\rangle \right) +  \rho_0 \partial_j \langle \tilde{v}_i \tilde{v}_j\rangle = - \partial_i \bar{p}  + \mu\partial^2_{jj} \bar{v}_i  + \left(\frac{\mu}{3}+\xi \right) \partial^2_{ij} \bar{v}_j, 
\label{eq:momentuma}
\end{equation}
since $\Pi_i = c_0^2\tilde{\rho}\tilde{v}_i$ at leading order. 

These two equations can be simplified to some extent with weakly restrictive hypotheses. First, if we consider the acoustic streaming produced by bulk acoustic wave (away from boundaries), the third term of the mass conservation equation (\ref{eq:massa}) is proportional at leading order to the bulk viscous dissipation of the wave energy, which remains weak in most media. This is quantified by the acoustical Reynolds number $Re_{ac}$ which compares the viscous dissipation to inertia  or equivalently the wave attenuation length $L_a = \rho c_0^3 / \omega^2 \mu \left( 4/3 + \frac{\xi}{\mu} \right) $ to the wavelength $\lambda$: 
$$
Re_{ac} = \frac{L_a}{\lambda} =  \frac{\rho_0 c_0^2}{\omega \mu \left( 4/3 + \frac{\xi}{\mu} \right)}
$$
Except at very high frequency ($> 1$ GHz) or for extremely viscous fluids and high driving frequencies, the acoustical Reynolds number is generally high ($Re_{ac} \gg 1$). The acoustical Reynolds number is estimated for the frequency and liquids used in the present experiments and simulations in table \ref{table1}.

Moreover if we consider only steady streaming (stationary flow produced by acoustic waves), the time derivatives in equations (\ref{eq:massa}) and (\ref{eq:momentuma}) can be canceled out. We obtain in this case:
\begin{equation}
\partial_i \bar{v}_i = 0
\label{eq:massaf}
\end{equation}
which amounts to saying that the steady streaming flow is incompressible. Then the time average momentum conservation equation becomes:
\begin{equation}
- \partial_i  \bar{p}  + \mu\partial^2_{jj} \bar{v}_i +  F_i = 0
\label{eq:momentumaf}
\end{equation}
with $F_i$ the Reynolds stress imbalance of the sound wave:
\begin{equation}
F_i = - \left\langle \rho_0\partial_j(\tilde{v}_i\tilde{v}_j)\right\rangle =  - \rho_0 \left\langle \tilde{v}_j \partial_j \tilde{v}_i + \tilde{v}_i \partial_j \tilde{v}_j\right\rangle . 
\label{eq:proof: Force term}
\end{equation}
This equation is simply the steady-state Stokes equation driven by a forcing term $F_i$ resulting from average nonlinear interactions of the acoustic field. 

It is worth noting that the derivation of acoustic streaming constitutive equations follows a similar procedure as the one used for the derivation of the Reynolds averaged Navier-Stokes equation in the field of turbulence. It describes how some strong fluctuating nonlinear terms influence the steady flow (\cite{Vanneste2011,Buhler2009}). Nevertheless, the fundamental differences between the derivation of the constitutive equations of acoustic streaming and turbulence are (i) that owing to the weak amplitude of the acoustic field, a perturbation analysis is possible, and (ii) that the source term in the average equations  emanates in the former case from the first order compressible field, namely the acoustic wave.

\subsection{Periodic fluctuations up to second order: nonlinear acoustics}
The mass and momentum equations for the periodic fluctuations $\tilde{f}$ up to second order can be obtained by subtracting the time-averaged equations (\ref{eq:massa}) and (\ref{eq:momentuma}) from the initial Navier-Stokes isentropic equations (\ref{eq:mass}) and (\ref{eq:momentum}):
\begin{equation}
\partial_t \tilde{\rho} + \rho_0 \partial_i \tilde{v}_i = - \partial_i \ll \tilde{\rho} \tilde{v_i} \gg ,
\label{eq:masswnl}
\end{equation}
and:
\begin{equation}
\rho_0 \partial_t \tilde{v}_i  + \partial_i \tilde{p} - \mu\partial^2_{jj} \tilde{v}_i - \left(\frac{\mu}{3}+\xi\right)\partial^2_{ij}\tilde{v}_j = - \partial_t \ll \tilde{\rho} \tilde{v_i} \gg - \rho_0 \partial_j \ll \tilde{v_i} \tilde{v_j} \gg,
\label{eq:momentumwnl}
\end{equation}
with $\ll \tilde{f} \tilde{g} \gg = \tilde{f} \tilde{g} - \left\langle \tilde{f} \tilde{g} \right\rangle$.

The left hand side of equations (\ref{eq:masswnl}) and (\ref{eq:momentumwnl}), along with the equation of state (\ref{eq:state}) at first order, constitute the linear equations of damped acoustic waves. The right hand side of these equations correspond to nonlinear terms, which modify the propagation of acoustic waves through energy transfers to harmonic frequencies ($2 \omega$, $3 \omega$, ...).

If we assume (see previous section) that the acoustical Reynolds and Mach numbers are small, these equations become at leading order:
\begin{eqnarray}
& & \partial_t \tilde{\rho} + \rho_0 \partial_i \tilde{v}_i = 0, \label{eq:dal1} \\
& & \rho_0 \partial_t \tilde{v}_i  + \partial_i \tilde{p}  = 0, \label{eq:dal2} \\
& & \mbox{with } \tilde{p} = c_0^2 \tilde{\rho} \label{eq:dal3},
\end{eqnarray}
which amounts to discard all nonlinear and dissipative effects.
From equation (\ref{eq:dal2}), we can infer that the oscillating flow is potential at leading order ($\tilde{v}_i = - \partial_i (\tilde{\phi}) $, with $\tilde{\phi}$ the velocity potential). A simple combination of equations (\ref{eq:dal1}) and (\ref{eq:dal3}): $\left[c_0^2 \; \partial_i \right.$(\ref{eq:dal1})$ -  \; \partial_t$(\ref{eq:dal2})$\left. \right]$ with (\ref{eq:dal3}) yields the celebrated d'Alembert equation:
\begin{equation}
\partial^2_{tt}   \tilde{\phi} - c_0^2 \partial^2_{ii} \tilde{\phi}  = 0 \label{eq:dalembert}
\end{equation}
with $\tilde{p} = \rho_0 \partial_t \tilde{\phi}$ and $\tilde{\rho} = \rho_0/c_0^2 \partial_t \tilde{\phi}$.

Now, if we do the same combination of equations (\ref{eq:masswnl}), (\ref{eq:momentumwnl}) and (\ref{eq:state}) but up to next order in $M$ and $1/Re_{ac}$, we obtain the Kuznetsov equation (\cite{spa_kuznetsov_1970}) (see e.g. \cite{ja_coulouvrat_1992} for a detailed demonstration with asymptotic analysis):
\begin{equation}
\partial^2_{tt}   \tilde{\phi} - c_0^2 \partial^2_{ii} \tilde{\phi}   - \frac{\mu b}{\rho_0}\partial_t \partial^2_{jj} \tilde{\phi} = 
\partial_t \left( \frac{B}{2A c_0^2}  \ll \left( \partial_t \tilde{\phi} \right)^2 \gg + \ll \left( \partial_i \tilde{\phi} \right)^2 \gg \right) \label{eq:kuznetsov_int}
\end{equation}
with $b = \frac{4}{3} + \frac{\xi}{\mu}$. Finally in the paraxial approximation (weak diffraction of the beam), we have the Lighthill-Westervelt equation$\left( \partial_i \tilde{\phi} \right)^2 = \frac{1}{c_0^2} \left( \partial_t \tilde{\phi} \right)^2$, leading to :
\begin{equation}
\partial^2_{tt}   \tilde{\phi} - c_0^2 \partial^2_{ii} \tilde{\phi}   - \frac{\mu b}{\rho_0}\partial_t \partial^2_{jj} \tilde{\phi} = 
\frac{\beta}{c_0^2} \partial_t  \ll \left( \partial_t \tilde{\phi} \right)^2 \gg 
\label{eq:kuznetsov}
\end{equation}
with $\beta = 1 + \frac{B}{2A}$ the so-called nonlinear parameter. This equation allows to compute the damped nonlinear propagation of acoustic waves.

The question then arises as to whether the nonlinear propagation of the acoustic wave must be considered to compute the acoustic streaming sources in equation (\ref{eq:momentumaf})? An elementary analysis solely based on the order of the nonlinear terms might lead to the misleading premature conclusion that since acoustic nonlinear terms on the right hand side of equation (\ref{eq:kuznetsov_int}) are of second order, their quadratic combination in equation (\ref{eq:proof: Force term}) is of fourth order and thus could be safely neglected when computing the acoustic streaming. In fact these nonlinear terms are weak but nevertheless cumulative. So they can play a significant role over a distance called the "shock distance" $L_s = \frac{c_0^2}{\omega \beta U_{ac}}$, which depends on the  first order velocity magnitude $U_{ac}$. To answer correctly to this question, it must thus be reminded that acoustic streaming is a consequence of the attenuation of the acoustic wave. This attenuation is proportional to the square of the acoustic wave frequency $\omega^2$. Since nonlinear terms in equation (\ref{eq:kuznetsov}) induce energy transfers from the driving frequency to higher harmonics, they promote the dissipation and thus the acoustic streaming. In an unbounded medium, the streaming enhancement by harmonics generation can be quantified by the ratio of the wave attenuation length $L_a = \frac{\rho_0 c_0^3}{\omega^2 \mu b}$ to the shock distance $L_s$:
$$
\frac{L_a}{L_s} = \frac{\rho_0 c_0 \beta U_{ac}}{\omega \mu b}
$$
Nonlinear terms in equation (\ref{eq:kuznetsov}) can thus be neglected when $L_a / L_s \ll 1$. In the present situation, since $U_{ac} < 10$ mm/s, the maximum value of this ratio is $10^{-1}$ for water and goes down to $10^{-3}$ for water-glycerol mixtures. Moreover, in cavities with water/air interfaces such as drops, the shock distance must also be compared to the size of the cavity $L_c$. Indeed, nonlinear effects are only significant when they are cumulative. Since it was shown by \cite{pre_tanter_2001} that each wave reflection at an air-water interface results in the deconstruction of these nonlinear effects, these latter can only be significant if the characteristic size of the cavity $L_c$ is larger than the shock distance. Here $L_c \sim 1$ mm while $L_s \sim 1$ m. Consequently, nonlinear terms can be safely discarded in equation (\ref{eq:kuznetsov})  for the analysis and simulation of the present experiments, leading to the equation of damped acoustic waves:
\begin{equation}
\partial^2_{tt}   \tilde{\phi} - c_0^2 \partial^2_{ii} \tilde{\phi}   - \frac{\mu b}{\rho_0}\partial_t \partial^2_{jj} \tilde{\phi} = 0
\label{eq:acousticlindamp}
\end{equation}

\subsection{Streaming source term: the hydrodynamic Reynolds stress tensor}
\label{sect: Hydrodynamic Reynolds stress tensor}
In this section we follow \cite{Eckart} and \cite{Vanneste2011} guidelines to analyze the different contributions of the streaming source (\ref{eq:proof: Force term}) and discard all the terms that do not actually produce acoustic streaming. This simplification is essential since the magnitude of the neglected terms is much larger than that of the relevant terms and can lead to significant numerical error  when calculating the flow produced by the acoustic wave.

Since the wave perturbation is irrotational, the first term of the force in equation (\ref{eq:proof: Force term}) is easily integrated and identified as the kinetic energy $\mathcal{K} = \frac{1}{2}\rho_0 \tilde{v}^2$:
\begin{equation}
\rho_0 \left\langle \tilde{v}_j \partial_j \tilde{v}_i \right\rangle = \partial_i\left\langle\mathcal{K}\right\rangle .
\label{eq:Fi1}
\end{equation}
The second term is computed from the mass conservation equation (\ref{eq:masswnl}) at leading order:
\begin{equation}
\left\langle \rho_0 \tilde{v}_i \partial_j \tilde{v}_j\right\rangle = -\left\langle \tilde{v}_i \partial_t\tilde{\rho}\right\rangle = \left\langle \tilde{\rho} \partial_t \tilde{v}_i \right\rangle ,
\end{equation}
where we used integration by part to move from the second expression to the third one. Then, if we use the classical vector identity $\mathbf{\nabla\nabla}\cdot\mathbf{\tilde{u}}=\mathbf{\nabla}^2\mathbf{\tilde{u}}+\mathbf{\nabla}\times\mathbf{\nabla}\times\mathbf{\tilde{u}}$, the wave momentum equation (\ref{eq:momentumwnl}) yields:
\begin{equation}
\left\langle \rho_0 \tilde{v}_i \partial_j \tilde{v}_j\right\rangle = -\left\langle \frac{\tilde{\rho}}{\rho_0} \partial_i \tilde{p}\right\rangle + \left\langle \frac{\tilde{\rho}}{\rho_0} \left(\frac{4}{3}\mu+\xi\right)\partial^2_{jj}\tilde{v}_i \right\rangle.
\label{eq:Fi2}
\end{equation}
Finally the equation of state (\ref{eq:state}) yields: $\langle\frac{\tilde{\rho}}{\rho_0}\partial_i\tilde{p}\rangle = \langle \frac{c^2}{2\rho_0}\partial_i\tilde{\rho}^2 \rangle$. Consequently, this term is the gradient of the potential energy of the wave in a linear medium:
\begin{equation}
\left\langle\frac{\tilde{\rho}}{\rho_0}\partial_i\tilde{p}\right\rangle = \partial_i\left\langle\mathcal{V}\right\rangle.
\label{eq:Fi3}
\end{equation}
If we combine equations (\ref{eq:Fi1}), (\ref{eq:Fi2}) and (\ref{eq:Fi3}), we obtain the following expression of the force $\mathcal{F}_i$ (see e.g. \cite{Gusev1979}):
\begin{equation}
\mathcal{F}_i = -\partial_i\left\langle\mathcal{L} \right\rangle - \left\langle \frac{\tilde{\rho}}{\rho_0} \left(\frac{4}{3}\mu+\xi\right)\partial^2_{jj}\tilde{v}_i \right\rangle ,
\label{eq:Fi}
\end{equation}
with the acoustic Lagrangian $\mathcal{L} = \mathcal{K} - \mathcal{V}$. We can work out a more practical equation by substituting the linear undamped wave equation in the viscous term, and assuming an harmonic wave motion:
\begin{equation}
\mathcal{F}_i = -\partial_i\left\langle\mathcal{L} \right\rangle +  \frac{\omega^2\nu b}{c^4} \left\langle \Pi_i \right\rangle,
\label{eq: complete force}
\end{equation}
with $\nu = \mu/\rho_0$. The first term derives from a potential and independent of the bulk and shear viscosities, and thus of the wave attenuation. Since acoustic streaming relies on the pseudo-momentum transfer from the wave mode (irrotational, compressible) to the viscous mode (solenoidal), through the wave attenuation, this term does not contribute to the steady flow; it is simply balanced by a hydrostatic pressure gradient. This can be easily verified by recasting equation (\ref{eq:momentumaf}) under the form:
\begin{equation}
- \partial_i  \bar{p^*}  + \mu\partial^2_{jj} \bar{v}_i +  \mathcal{F}_i^* = 0
\label{eq:momentumaff}
\end{equation}
with $\bar{p}^* = \bar{p} + \left\langle\mathcal{L} \right\rangle$  and $\mathcal{F}_i^* =  \frac{\omega^2\nu b}{c^4} \left\langle \Pi_i \right\rangle$. $\mathcal{F}_i^*$ is related to the wave dissipation has a much smaller magnitude than its counterpart $\left\langle\mathcal{L} \right\rangle$. Nevertheless, it is not potential and hence is the sole source term of acoustic bulk streaming as emphasized in \cite{Lighthill1978391} and \cite{Eckart} studies.

\subsection{Final simplified system of equations}

In this section, we derived the constitutive equations of acoustic streaming and optimized the expressions for the simulation of the experimentally observed acoustic streaming in droplets:
\begin{eqnarray}
& & \mbox{Acoustic wave: } \partial^2_{tt}   \tilde{\phi} - c_0^2 \partial^2_{ii} \tilde{\phi}   - \frac{\mu b}{\rho_0}\partial_t \partial^2_{jj} \tilde{\phi} = 0 \label{feq:mass} \\
& & \mbox{Streaming Stokes flow: } \left\{ \begin{array}{c}  \partial_i{\bar{v}_i} = 0 \\ \\ - \partial_i  \bar{p^*}  + \mu\partial^2_{jj} \bar{v}_i +  \mathcal{F}_i^* = 0  \end{array} \right. \label{eqf:streaming} \\
& & \mbox{Acoustic streaming source term: } \mathcal{F}_i^* =  \frac{\omega^2\nu b}{c^4} \left\langle \Pi_i \right\rangle. \label{eqf:source}
\end{eqnarray}
We also highlighted some similitudes between the derivation of the constitutive equations of acoustic streaming and turbulence. Indeed, acoustic streaming arises from the imbalance of Reynolds stress like turbulence, but acoustics allows an exact computation of the forcing term. This analogy will be used in the next section to develop the equivalent of the Large Eddy Simulation numerical method in the field of acoustic that will be referred in the following as the Streaming Sources Spatial Filtering (SSSF) method.

\section{Numerical model} \label{sec: Numerical model}
\begin{figure} \begin{center}
\includegraphics{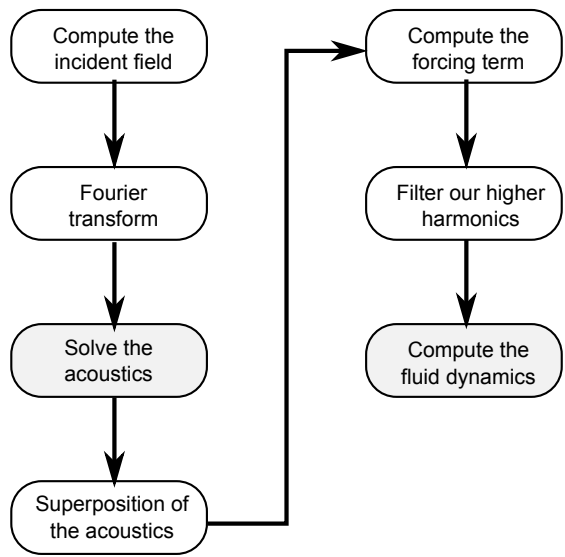}
\caption{Computational method flowchart. White steps were performed with Matlab and grey steps with Comsol.}
\label{fig: comput flowchart}
\end{center} \end{figure}

As stated in the experimental section, when a highly viscous droplet is exposed to megahertz surface acoustic wave excitation, the hydrodynamic flow may take up to tens of seconds to reach steady state. This colossal difference of time scales between acoustics and hydrodynamics prevents any attempt to compute dynamically the acoustic streaming in complex 3D geometries. Instead, we first simulate the acoustic field and then the hydrodynamic flow, as shown in figure \ref{fig: comput flowchart}. The computation is not  straightforward due to the large discrepancy between the droplet size and the acoustic wavelength. The acoustic problem is solved in cylindrical geometry to minimize memory usage, the incident field being resolved as a sum of cylindrical functions by spatial Fourier transform. This allows to compute the three-dimensional acoustic field through several elementary calculations of acoustic fields in a two-dimensional geometry with reduced memory requirements. Then, the hydrodynamic flow is computed with a simplified forcing term reminiscent of the Large Eddy Simulation (LES) which mimics the effect of viscosity to minimize complex momentum source terms.

\subsection{Computation of the acoustic field}
The acoustic field is computed in the frequency domain. In this case, equation (\ref{feq:mass}) becomes:
\begin{equation}
\partial^2_{ii} \tilde{\phi} + k^2 {\tilde{\phi}} = 0
\label{eq: Helmoltz Cartesian}
\end{equation}
with: 
\begin{equation}
k^2 =\frac{k_0^2}{1+i / Re_{ac}} .
\end{equation}
Here, $k_0 = \omega/c_0$ is the wavenumber of the unattenuated wave and $Re_{ac}$ is the acoustical Reynolds number ($Re_{ac} \gg 1$). 

The large discrepancy between the droplet size and the acoustic wavelength yields very large and intensive simulations. For instance, direct 3D simulation of the acoustic field on a 32 GB RAM computer with the finite element method only allows to simulate 2 mm diameter droplets up to 8 MHz. As shown in figure \ref{fig: memory DNS SSSF} in the appendix, RAM requirements sharply increase with increasing frequency, and extrapolation to 20 MHz culminates at 1.0 TB, thus preventing any direct computation of the acoustic field.

To minimize memory requirements, we took advantage of the droplet rotational symmetry to reduce the problem to a two-dimensional one. The protocol described in the following uses Fourier transform to resolve the incident field as a sum of circular harmonics, solve each of them separately and then reconstruct the field thanks to the superposition principle. In this way, the complete problem is decomposed into sub-problems with low memory requirements which can be computed in parallel. This method thus ensures an optimal matching with the capacity of the computer (number of cores, memory).
\subsubsection{Method: Spatial Fourier Transform \label{sss:msft}} 
\begin{figure} \begin{center}
\includegraphics{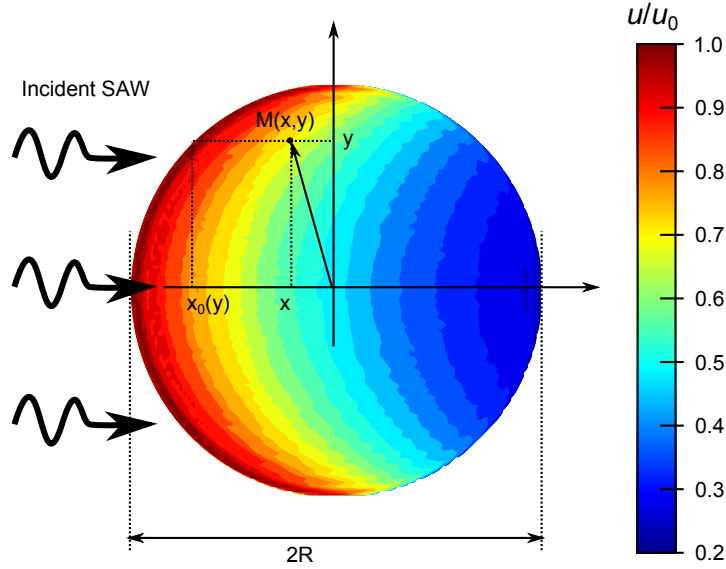}
\caption{Model of the incident leaky SAW. Color are indicative of the SAW magnitude. At 20 MHz and for a 12.5 $\mu$L water droplet (2R = 3.7 mm), the incident SAW vertical displacement $\tilde{u}$ drops by 80\% as it propagates beneath the droplet.}
\label{fig: leaky SAW drawing}
\end{center} \end{figure}

Working in cylindrical coordinates, Fourier transform allows resolving any function into a convenient weighted sum of complex exponentials: 
\begin{equation}
f(r,\theta) = \sum_{l = -\infty}^{+\infty} f_l(r) e^{i l \theta},
\end{equation}
with
\begin{equation}
f_l(r) = \frac{1}{2\pi}\int_{-\pi}^{+\pi} f(r,\theta) e^{- i l \theta} d\theta .
\end{equation}
Here, the only non-axisymmetric boundary condition is the normal displacement $\tilde{u}$ of the substrate due to the incident SAW. It is projected into Fourier harmonics:
\begin{equation}
\tilde{u}_l(r) = \frac{1}{2\pi}\int_{-\pi}^{+\pi} \tilde{u}(r,\theta) e^{- i l \theta} d\theta .
\label{eq: incident wave Fourier}
\end{equation}
In practice, the value of $l$ can be restricted. Indeed, $l_{\mathtt{max}} \simeq \pi D/\lambda_s$ corresponds to the maximum number of wavelength $\lambda_s$ the input surface acoustic wave can travel along the perimeter of the droplet, where $D$ is the droplet diameter. We computed this integral for $l\in [ 0,2l_\mathtt{max} ]$, $l$ being an integer. The value of $\tilde{u}$ depends on the incident wave. Neglecting diffraction, the SAW magnitude decreases exponentially as soon as it meets the droplet interface at a given point $x_0$. The attenuation rate $\alpha$ is provided for instance by \cite{CampbellJones}. For a given point $M(x,y)$, the propagation length beneath the droplet is given by $x-x_0(y)$ (see figure \ref{fig: leaky SAW drawing}). The vertical displacement field $\tilde{u}$ at the droplet base is then given by \cite{jjap_shiokawa_1990}:
\begin{eqnarray}
\tilde{u} & = & u_0 \exp(-i k_s x) \exp(-\alpha(x-x_0(y))) \label{eq: incident wave1}\\
x_0(y) & = & -\sqrt{R^2-y^2}\\
\alpha & = & \alpha_0 \ln(10) F/20,
\label{eq: incident wave3}
\end{eqnarray}
where $F$ is the SAW frequency in Hz, $u_0$ is the magnitude of the acoustic perturbation displacement and $\alpha_0$ the attenuation coefficient (in s/m)  $\alpha_0\simeq 2.0 . 10^{-7} \times \rho_0$. We computed $\alpha_0$ from the value given by \cite{CampbellJones} in the case of water loading (0.2 dB/MHz/cm).

We then solve all variables in the form: $ \tilde{p}(r,\theta,z)= \tilde{p}_l(r,z) e^{i l \theta}$, $\tilde{v}_{j}(r,\theta,z)= \tilde{v}_{j,l}(r,z) e^{i l \theta}$ with $j$ either $r,\theta,z$.
Equation (\ref{eq: Helmoltz Cartesian}) becomes:
\begin{equation}
k^2 \tilde{p}= - \frac{1}{r} \partial_r \left(r \partial_r \tilde{p}\right)  - \frac{1}{r^2}\partial^2_{\theta\theta} \tilde{p}-\partial^2_{zz} \tilde{p},
\label{eq: Helmoltz Polar}
\end{equation}
which can be re-casted using the axisymmetric variables:
\begin{equation}
\left(k^2 - \frac{l^2}{r^2}\right) \tilde{p}_l= - \partial^2_{rr} \tilde{p}_l   -\partial^2_{zz} \tilde{p}_l - \frac{1}{r}\partial_r \tilde{p}_l .
\label{eq: Helmoltz Polar2}
\end{equation}
This equation is solved with a finite element method by the commercial solver COMSOL 4.3b. \footnote{The default PDE interface for axisymmetric systems does not include the last right hand term of equation (\ref{eq: Helmoltz Polar2})}. The boundary condition at the liquid air interface reads:
\begin{equation}
\tilde{p}_l = 0 .
\label{eq: air-liq acoust BC} 
\end{equation}
At the solid liquid interface, we enforce an impedance boundary condition with the source term computed from equation (\ref{eq: incident wave Fourier}): 
\begin{equation}
\partial_z \tilde{p}_l = \rho_0\omega^2\tilde{u}_l - i \frac{\omega \rho_0}{\mathcal{Z}_s}\tilde{p}_l  ,
\label{eq: sol-liq acoust BC}
\end{equation}
where $\mathcal{Z}_s = \rho_s c_s$ is the acoustic impedance of the solid \footnote{This expression is exact only for plane waves with normal incidence. Nevertheless, the agreement of our simulations with experimental results was equally good at low attenuation (where this boundary condition might have played a role) and at higher viscosity (where this boundary condition is irrelevant since the wave is attenuated even before bouncing back to the solid)}. The solution $\tilde{p}$ can then be reconstructed thanks to the linearity of the equations:
\begin{equation}
\tilde{p}(r,\theta,z) = \sum_{-\infty}^{+\infty} \tilde{p}_l e^{i l \theta} .
\end{equation}
The velocity field can be reconstructed in a similar fashion:
\begin{equation}
\tilde{v}_r(r,\theta,z) = \frac {-1}{\rho_0 i\omega}\partial_r \tilde{p} =  \sum_{-\infty}^{+\infty} \tilde{v}_{r,l} e^{i l \theta} ,
\end{equation}
\begin{equation}
\tilde{v}_\theta(r,\theta,z) = \frac {-1}{\rho_0 i r \omega}\partial_\theta \tilde{p} =  \sum_{-\infty}^{+\infty} \tilde{v}_{\theta,l} e^{i l \theta} ,
\end{equation}
\begin{equation}
\tilde{v}_z(r,\theta,z) = \frac {-1}{\rho_0 i\omega}\partial_z \tilde{p} =  \sum_{-\infty}^{+\infty} \tilde{v}_{z,l} e^{i l \theta} ,
\end{equation}
with:
\begin{eqnarray}
\tilde{v}_{r,l} & = & \frac {-1}{\rho_0 i\omega}\partial_r \tilde{p}_l ,\\
\tilde{v}_{\theta,l} & = & \frac {-l}{\rho_0 \omega} \tilde{p}_l ,\\
\tilde{v}_{z,l} & = & \frac {-1}{\rho_0 i\omega}\partial_z \tilde{p}_l .
\end{eqnarray}
If the incident field is symmetric along the x-axis, we have:
\begin{equation}
f(r,\theta,z) = f_0+2\sum_{1}^{+\infty} f_l \cos(l \theta),
\label{eq: reconstruct even}
\end{equation}
where $f$ stands for either $\tilde{p}$, $\tilde{v}_r$ or $\tilde{v}_z$. $\tilde{v}_{\theta,l}$ is odd due to the factor $l$: 
\begin{equation}
\tilde{v}_\theta(r,\theta,z) =  2i\sum_{1}^{+\infty} \tilde{v}_{\theta,l} \sin(l \theta)
\label{eq: reconstruct odd}
\end{equation}

Thus, the acoustic field calculation can be summed up as follows: we start by computing the incident wave using equation (\ref{eq: incident wave1}-\ref{eq: incident wave3}), then its Fourier transform as given by equation (\ref{eq: incident wave Fourier}). We combine these data and boundary conditions (\ref{eq: air-liq acoust BC}, \ref{eq: sol-liq acoust BC}) with the wave equation (\ref{eq: Helmoltz Polar2}) to obtain the acoustic field for each individual harmonic. Finally, the total field is reconstructed using equations (\ref{eq: reconstruct even}, \ref{eq: reconstruct odd}). The algorithms ensuring the azimuthal Fourier transform were checked carefully by comparing the acoustic field in droplets exposed to 6 MHz as computed by a direct finite element model and by the Fourier method (figures available in supplemental material).

\subsubsection{Resulting acoustical field in the droplet} \label{sec: Acoust field in the droplet}
To the best of our knowledge, the current computation of the acoustic field in a 3D sessile droplet involves frequencies an order of magnitude above the only other work published so far by \cite{Quintero2013}. As a result, it is significantly different and we will dedicate a few lines to detail the key features of this field.

\begin{figure} \begin{center}
\includegraphics[width=0.5 \textwidth]{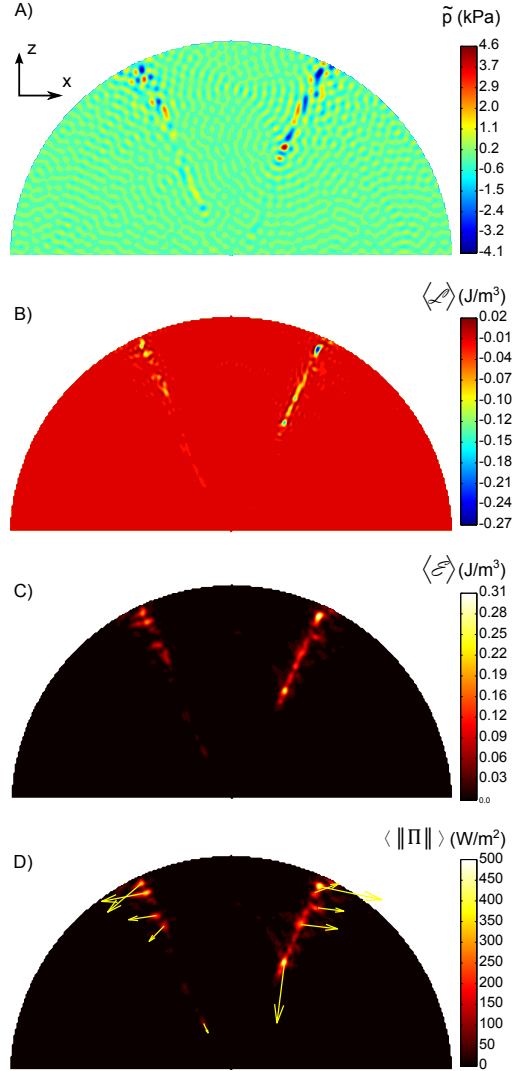}
\caption{Meridian cross-section of the acoustic field in a water droplet excited by a 20 MHz acoustic field. A) Acoustic pressure $\tilde{p}$. $\tilde{p}_{max} = 40$ kPa B) Average Langrangian density $\langle \mathcal{L} \rangle$. $-0.27$ J $ < \langle \mathcal{L} \rangle < 0.020$ J/m$^3$. C) Average energy density $\langle\mathcal{E}\rangle=\langle\mathcal{K}+\mathcal{V}\rangle$. $\langle \mathcal{E}\rangle_{max} = 0.31$ J/m$^3$. D) Poynting vector. $\| \langle \mathbf{\Pi} \rangle \|_{max} = 300$ W/m$^2$. Droplet volume is 12.5 $\mu$L, base diameter is 3.7 mm. The incident wave comes from the left with a vertical displacement of 10 pm.}
\label{fig: acoustic field in the water droplet}
\end{center} \end{figure}

\begin{figure} \begin{center}
\includegraphics[width=0.5 \textwidth]{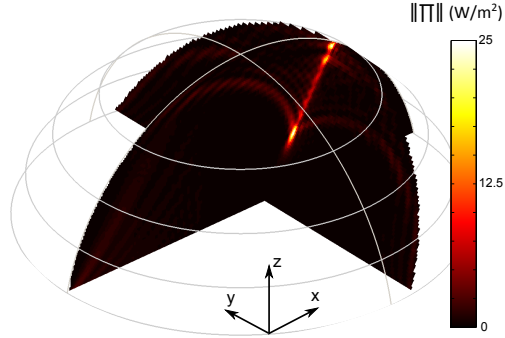}
\caption{Poynting vector in a 90 wt$\%$ glycerol droplet excited by a 20 MHz acoustic field. $\| \langle \mathbf{\Pi} \rangle \|_{max} = 25$ W/m$^2$ Droplet volume is 12.5 $\mu$L, base diameter is 4.0 mm. The incident wave propagates along the $x$ axis (from left to right) with an amplitude of 10 pm.}
\label{fig: poynting vector in the glycerol droplet}
\end{center} \end{figure}

In figure \ref{fig: acoustic field in the water droplet}, we show the acoustic field in a sessile water droplet excited by a 20 MHz SAW. The acoustic pressure (\ref{fig: acoustic field in the water droplet}.A) appears with two caustics superimposed on a quasi-random background field. The incident wave is overwhelmed by the numerous reflections on the droplet surface. The two caustics are much more pronounced than what is found in the two-dimensional analog \cite{pre_brunet_2010}, probably due to the increased ray convergence in 3D. The Lagrangian of the acoustic field (\ref{fig: acoustic field in the water droplet}.B) is mostly focused along the two caustics. In models based on Nyborg's force in the continuity of Shiokawa \textit{et al.} , the gardient of this Lagrangian is used as the driving force of acoustic streaming. Since its expression derives from a gradient, it is similar to a potential energy and therefore it can be exactly compensated by a hydrostatic pressure term. The energy of the acoustic field (\ref{fig: acoustic field in the water droplet}.C) clearly shows the predominance of the caustics. The angle of this concentration of energy matches the Rayleigh refraction angle, and the symmetry shows the stability of this particular acoustic ray. Finally, the Poynting vector (\ref{fig: acoustic field in the water droplet}.D) is similar to the energy density and the Lagrangian, except that it is almost divergenceless in weakly attenuating media, and it gives some directions for the flow forcing that are consistent with the experiment. Importantly, the power flow is concentrated in the meridian plane up to the diffraction limit, as shown in figure \ref{fig: poynting vector in the glycerol droplet}.

\begin{figure} \begin{center}
\includegraphics[width=0.5 \textwidth]{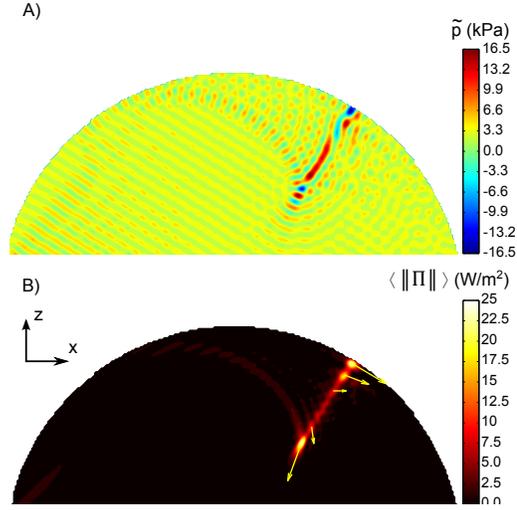}
\caption{Meridian cross-section of the acoustic field in a 90 wt$\%$ glycerol droplet excited by a 20 MHz acoustic field. A) Acoustic pressure $\tilde{p}_{max} = 16.5$ kPa. B) Poynting vector. $\| \langle \mathbf{\Pi} \rangle \|_{max} = 25$ W/m$^2$ Droplet volume is 12.5 $\mu$L, base diameter is 4.0 mm. The incident wave comes from the left with an amplitude of 10 pm.}
\label{fig: acoustic field in the glycerol droplet}
\end{center} \end{figure}

In figure \ref{fig: acoustic field in the glycerol droplet}, we display the radiation patterns obtained in 90 wt$\%$ glycerol droplets. At higher viscosity, the pressure field becomes less symmetrical. Indeed, the incident wave is attenuated faster and hence undergoes less reflections at the droplet surface. The Rayleigh radiation angle appears more clearly at higher dissipation, and the wave pattern looses its left-right symmetry. Remarkably, the Poynting vector becomes completely asymmetrical and forces the flow on a single side of the droplet. We will detail the consequences of this change on the resulting flow pattern in the next section.

\begin{figure} \begin{center}
\includegraphics{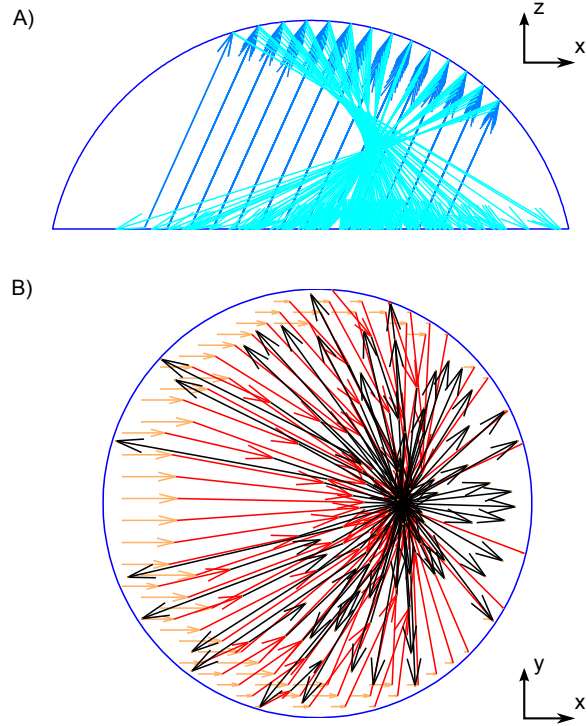}
\caption{Geometrical acoustics interpretation of the caustics. A) Side view of the primary focus formed by the direct reflection of the incident beam. B) Top view of the secondary focus formed by 3rd degree reflections on the droplet interface and showing a significant momentum imbalance.}
\label{fig: caustics geometrical interpretation}
\end{center} \end{figure}

Since the caustics play a major role in driving the droplet internal flow, we traced back their origin in figure \ref{fig: caustics geometrical interpretation}. We distinguish between a primary focus (\ref{fig: caustics geometrical interpretation}.A) formed by the direct reflection of incident beams on the concave droplet interface and a secondary one (\ref{fig: caustics geometrical interpretation}.B) formed by third degree reflections on the droplet interface. For some reason, secondary reflections do not form focal points. The first focus shows a strong asymmetry along the $z$ direction while the second focus is not symmetrical along the $x$ direction. Since the arrows represent rays, which are related to the Poynting vector, and since the Poynting vector is the forcing term of acoustic streaming, the asymmetry indicates some net momentum influx.

\subsection{Computation of the resulting flow} \label{sec: DNS_vs_SSSF}

\subsubsection{Direct numerical simulation (DNS)}

The flow is computed with equations (\ref{eqf:streaming}) and (\ref{eqf:source}). These equations are combined with the no-slip boundary conditions at the solid-liquid interface and shear-free boundary condition at the air-liquid interface to perform the direct numerical simulation (DNS) of droplet acoustic streaming. DNS is a simulation from first principles and easy to implement.

The major shortcoming of DNS is the extensive use of memory. Indeed, the Poynting vector is a second-order quantity and has a typical variation length-scale of $\lambda/2$. Consequently, DNS becomes computationally prohibitive for frequencies above 6 MHz. In the appendix, we report the memory required to simulate droplets exposed to SAW radiations with frequencies up to 8 MHz. Extrapolation to 20 MHz indicates that up to 580 GB of RAM would be necessary to perform the DNS simulation of our experiments.

\subsubsection{Streaming Source Spatial Filtering (SSSF)}

In the world of turbulence, the gigantic difference of length-scale between the main flow patterns and the smallest eddies resulting from the break up of large flow structure is a major issue. A well-established method called Large Eddy Simulation (LES) allows computation of turbulence on relatively coarse grids that account for Sub-Grid Scale (SGS) dynamics through an SGS model (see e.g.\cite{Deardorff1970,Pope2004,Bou-Zeid2015}). The case of acoustic streaming appears as a reverse situation wherein a large scale flow (the acoustic streaming) emerges from small scale fluctuations (the acoustic wave). The Streaming Source Spatial Filtering (SSSF) method presented in this section  relies on the fact that the small scales variations of the streaming source term $\mathcal{F}^*$ do not contribute to the flow since they are filtered by the fluid viscosity. Indeed, the acoustic streaming under investigation is slow and laminar, yielding a linear equation with a momentum source term. It is then interesting to consider the velocity field in the reciprocal space in terms of spatial harmonics. We can match each wavenumber of the velocity field with a (possibly null) forcing term in order to solve each equation independently. It is then straightforward to notice that higher wavenumbers are filtered out by the Laplacian operator of the viscosity (decreasing in $1/k^2$).  In this regard, this SSSF model differs significantly from LES: in the former, smaller scales are \textit{sources} of momentum, and dissipation happens at larger scales, whereas in the latter smaller scales acts a momentum \textit{sinks} because large-scale flows dissipate little energy.

Such filtering enables to use grid cell sizes for the flow computation larger than the acoustic wavelength and thus considerably reduces the computational requirements for the resolution of the flow problem. It is important to note that since we work in the small Reynolds number regime, the characteristic length scale of the flow is entirely dictated by the streaming source term and the boundary conditions (no additional scale emerges from the flow itself like in the case of turbulent flows). The filtered source term $\underline{\mathcal{F}^*}$ is obtained in the real space from the convolution product with the filtering function $\mathcal{H}(x,y,z)$:
\begin{equation}
\underline{\mathcal{F}^*}= f\ast \mathcal{H} ,
\end{equation}
where the filtering function $\mathcal{H}(x,y,z)$ is defined from the filter transfer function $H(k_x,k_y,k_z)$ according to the formula:
\begin{equation}
\mathcal{H}(x,y,z) =\int_{\mathcal{S}} H(k_x,k_y,k_z) e^{i k_x x + i k_y y + i k_z z} d\mathcal{S} ,
\label{eq: SSSF filter}
\end{equation} 
with $\mathcal{S}$ the reciprocal space, $H(k_x,k_y,k_z)=1$ when $k_x^2+k_y^2+k_z^2 < k_c^2$ and zero otherwise, and $k_c$ is the critical wavenumber of the filtered flow structures. We choose the critical wavenumber $k_c$ as half the acoustical wavenumber in the fluid at working frequency. Indeed, the acoustic forcing term is the product of two acoustic quantities, which halves the spatial period. The exact choice of $k_c = k/2$  is somewhat arbitrary provided $k_c$ is below $2k$ and larger than $2\pi/L$, where $L$ is the typical scale of the feature to be observed. This assertion was validated for a range of parameters, and results are provided in supplemental material.

Consequently, the equations solved with the SSSF method are simply:
\begin{eqnarray}
& &\partial_{i} \bar{v}_i = 0, \\
& &  - \partial_i \bar{p^*} + \mu\partial^2_{jj} \bar{v}_i + \underline{\mathcal{F}_i^*} = 0 .
\end{eqnarray}
In the next section, we will show that the flow patterns computed from the DNS and the SSSF method agree qualitatively and quantitatively. Hence, in the remaining part of the paper all the simulations at 20 MHz will be performed with the SSSF method to overcome hardware limitations.

\subsection{Results}
The results of the simulation are presented from the most technical aspects to the most physical ones. First we compare the flow pattern as given by the Direct Numerical Simulations and the Streaming Source spatial Filtering method, and then we show the physical results relevant to the experimental study. This section is supplemented by appendix \ref{sec: comput_cost} where we expose the memory requirements of direct methods versus the numerical recipes introduced previously.

\subsubsection{Comparison of DNS and SSSF}

Streaming Source Spatial Filtering and Direct Numerical Simulation are two very different numerical computational methods. In this section, we provide some insights on how and how fast the SSSF converges. First, we compare the results of a 2 mm diameter water sessile droplet exposed to 6 MHz SAW radiations with an amplitude of 10 nm.

\begin{figure} \begin{center}
\includegraphics{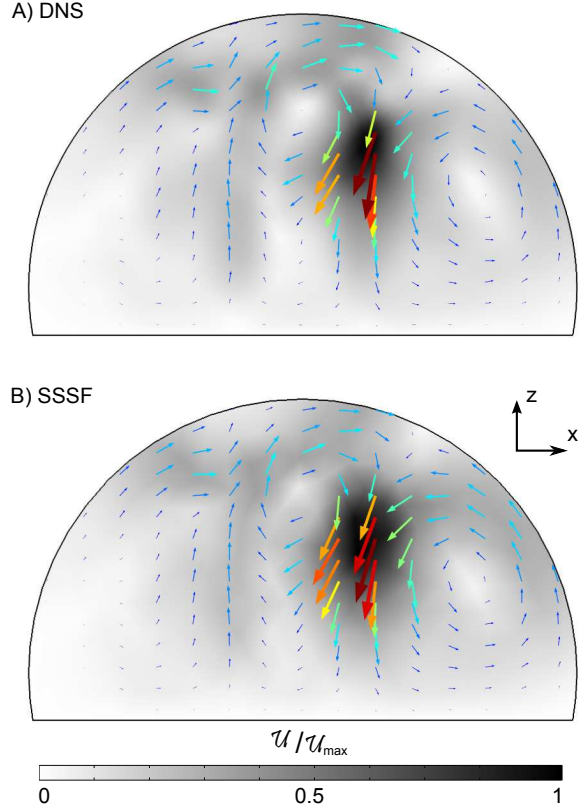}
\caption{Meridian cross-section of the hydrodynamic flow pattern in a water droplet excited by a 6 MHz acoustic field. A) Direct Numerical Simulation (DNS) on a fine grid (311,000 elements). B) Simulation with the SSSF method on a much coarser grid (11,090 elements). The velocity magnitude is indicated in gray scale, darker grays represent larger velocities. The flow patterns agree qualitatively well despite some mismatch on the velocity magnitude ($\mathcal{U}_{max} = 30$ mm/s for DNS versus 19 mm/s for SSSF). This discrepancy is discussed later in the text. Droplet contact angle $\theta_c$ is 100$^o$, and its base radius is 0.98 mm ($\alpha D = 0.27$, $\Lambda = 0.004$).}
\label{fig: comparison DNS SSSF}
\end{center} \end{figure}

\begin{figure} \begin{center}
\includegraphics{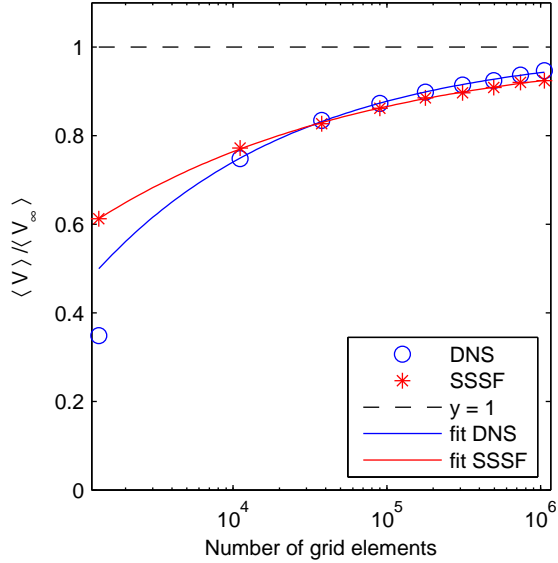}
\caption{Comparison of the convergence speed of Direct Numerical Simulation and Streaming Source Spatial Filtering methods versus the number of grid elements. The average flow speed $\langle V \rangle$ returned by the numerical simulations were normalized by an estimation of their asymptotic value $\langle V_{\infty} \rangle$. Nonlinear curve fitting returns $\langle V_{DNS} \rangle $(mm/s)$\simeq  3.1 - 16.4N^{-0.33}$ and $\langle V_{SSSF} \rangle $(mm/s)$ \simeq  2.6 - 5.8N^{-0.25}$. The simulation point computed from the coarsest grid was excluded during the fitting of the DNS since it appears no asymptotic convergence regime was reached at this point.}
\label{fig: convergence SSSF}
\end{center} \end{figure}

In figure \ref{fig: comparison DNS SSSF}, we display the flow fields computed by DNS and SSSF in the meridian plane of the droplet. The flow patterns are very similar, despite the large difference of memory requirements (2 GB for the SSSF, 10 GB for the DNS). This illustrates the ability of SSSF to show a convergent behavior even for very rough grids. In order to be more quantitative, we computed the average flow speed in the droplet with DNS and SSSF for 9 degrees of grid refinements (minimum element size $l=\lambda/n$, $n\in \{1..9\}$). Results shown in figure \ref{fig: convergence SSSF} were analyzed by nonlinear curve fitting. Accordingly, the DNS (SSSF) converges towards $\langle V_\infty \rangle = 3.1$ mm/s ($\langle V_\infty \rangle = 2.6$ mm/s) at a rate in $O(N^{1/3})$ ($O(N^{1/4})$) \footnote{We attribute the slight difference of asymptotic value of $\langle V \rangle$ (approximately 17\%) between DNS and SSSF to minor differences (8\%) in the acoustic field depending whether they were computed by DNS or circular Fourier transform decomposition as described in section \ref{sss:msft}.}. Importantly, DNS starts converging only after the number of grid elements has exceeded 10,000 as testified by the outlier on the left of the graph in figure \ref{fig: convergence SSSF}, whereas SSSF consistently follows its convergence trend even for a such a rough grid.


\subsubsection{Comparison with experiments}

\begin{figure*} \begin{center}
\includegraphics[width=\textwidth]{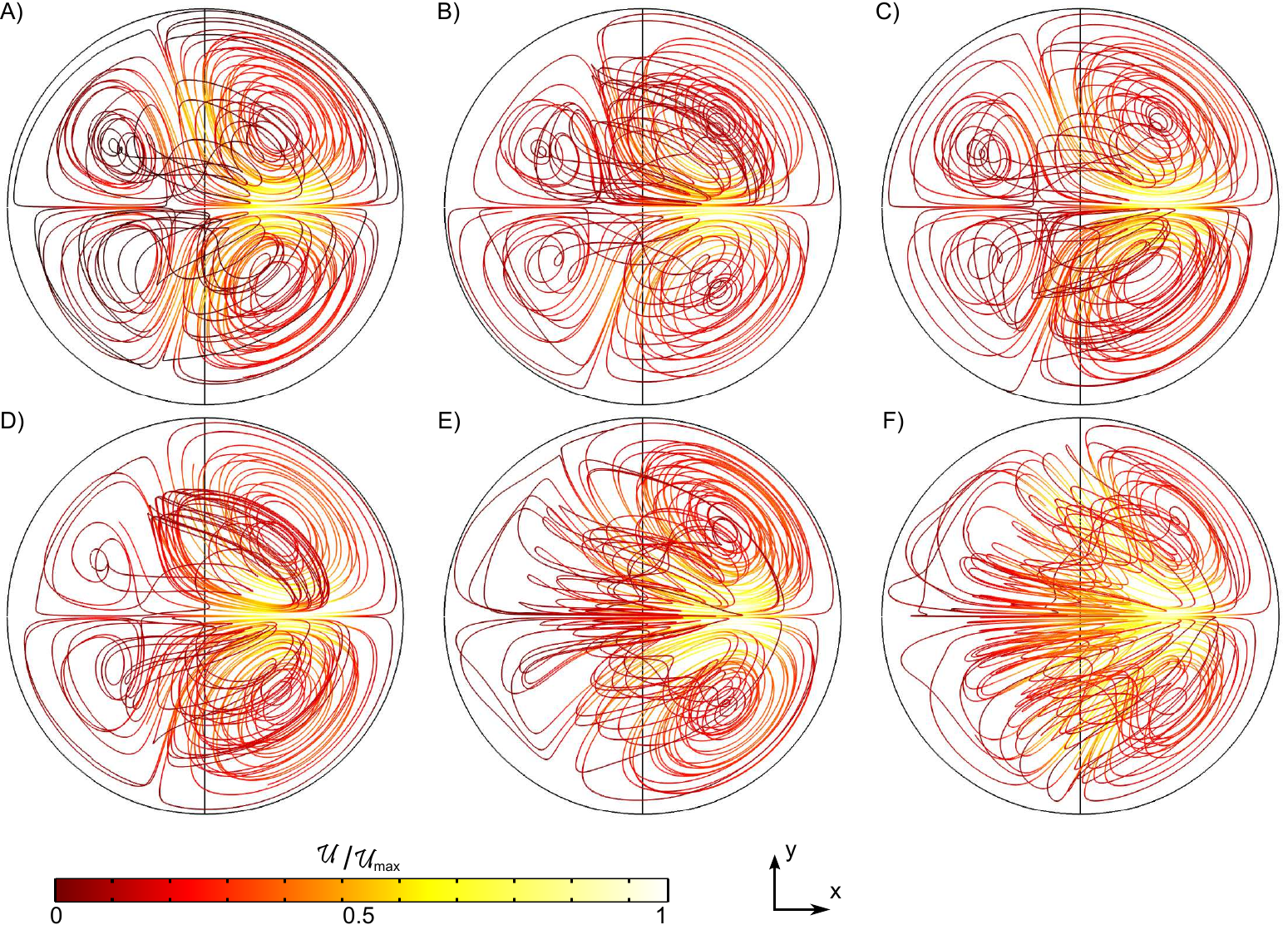}
\caption{Streamlines from SSSF flow field computations for the flow field for various glycerol concentrations. The visualization is from below. The SAW propagates from left to right. The droplet volume is $V_{droplet}$ = 12.5 $\mu$l and the magnitude of the acoustic perturbation displacement $u_0 = 44$ pm. As the viscosity increases, one remarks the progressive transition from a four-vortex to a two-vortex flow structure. (A) Pure Water ($\mathcal{U}_{max} = 173$ $\mu$m/s) (B) 30 wt$\%$ Glyc. ($\mathcal{U}_{max} = 170$ $\mu$m/s) (C) 40 wt$\%$ Glyc. ($\mathcal{U}_{max} = 137$ $\mu$m/s) (D) 60 wt$\%$ Glyc. ($\mathcal{U}_{max} = 61$ $\mu$m/s) (E) 80 wt$\%$ Glyc. ($\mathcal{U}_{max} = 23$ $\mu$m/s) (F) 90 wt$\%$ Glyc. ($\mathcal{U}_{max} = 3.8$ $\mu$m/s)}
\label{fig: droplet flow versus visc}
\end{center} \end{figure*}

In the previous sections, we have developed and characterized a numerical algorithm to compute the acoustic streaming in large cavities compared to the wavelength. We now apply it to sessile droplets  of various viscosities exposed to 20 MHz SAW radiations and compare it to experiments presented in section \ref{sec: Experiment}. Results are shown in figure \ref{fig: droplet flow versus visc}.

\begin{figure} \begin{center}
\includegraphics{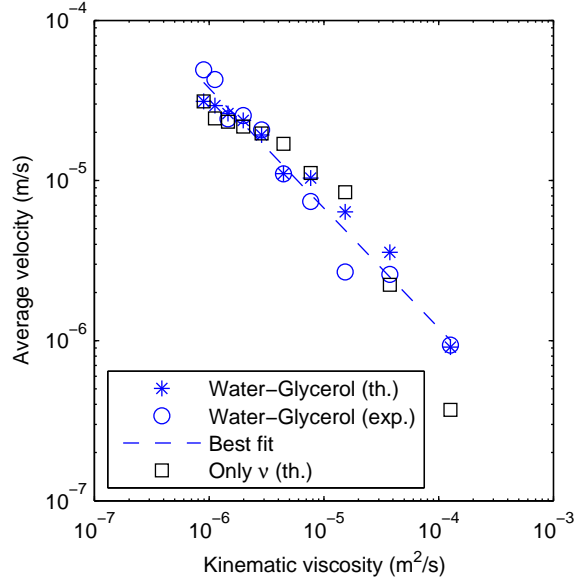}
\caption{Averaged velocity of the droplet internal flow versus viscosity. The \textit{stars} correspond to numerical simulations performed with water-glycerol physical properties summarized in table \ref{tab: physical parameters}, the \textit{circles} to experimental data, the \textit{squares} to numerical simulations performed by varying only the viscosity of the water-glycerol mixture and keeping all other parameters equal to those of water, and (4) the best fit with a power law ($\langle V \rangle \propto \nu^{-3/4}$). The only fitting parameter is the SAW amplitude, which is estimated to be 44 pm according to the simulations compared to 62 pm given by our measurements with the laser interferometer.}
\label{fig: quantitative flow speed}
\end{center} \end{figure}

Similarly to figure \ref{fig: experimental_top_view}, the droplet flow pattern progressively switches from four eddies to two eddies. The agreement is not only qualitative but also quantitative as shown in figure \ref{fig: quantitative flow speed} where we plot the average flow speed in the droplet versus the liquid viscosity. In this curve, the adjustable parameter was the solid displacement magnitude. Linear regression gives 44 pm which compares well to the 62 pm measured with a Doppler-shift interferometer (SH130, B.M. Industries). In order to segregate viscosity as the dominant factor for the change of velocity, we compare the experiment to two scenarios. In the first one, we implement all real values of physical quantities in the numerical model for the different water-glycerol mixtures (table \ref{tab: physical parameters}) whereas the second one keeps the physical properties of water for all quantities (contact-angle, density, sound speed) except for the viscosity which is set to the real water-glycerol system, yielding to idealized comparative situations where only the viscosity varies. The excellent agreement between the experiment and the realistic simulation validates our numerical model. More importantly, the good agreement between the idealized model and the experiments evidence unambiguously the strong dependence of Eckart acoustic streaming on the fluid viscosity.

\section{Discussion} \label{sec: DimensionlessAnalysis}
In the previous sections, we have presented the Eckart bulk streaming as the motile force of the flow observed in sessile droplets exposed to SAW radiations. We have developed a numerical model based on first principles to compute the acoustic streaming in three dimensions, and the results agree remarkably well with experimental data. In both cases, the flow pattern in the droplet shows a gradual transition from four to two eddies which has not been reported nor explained so far in the literature. In this section, we discuss these results based on our numerical model. Indeed, it unveils the acoustic field which generates the forcing term of the flow. This allows a qualitative and quantitative analysis of the flow development and enables to single out the most influential parameters, which are the caustics and the surface wave attenuation.

\subsection{Acoustic Forcing term and Flow pattern}

The acoustic forcing term, given by equation (\ref{eqf:source}) is proportional to the Poynting vector. This means that the knowledge of the acoustic power flow is tantamount to the knowledge of the forcing term. The Poynting vector in water and glycerol droplet is shown in figures \ref{fig: acoustic field in the water droplet} and \ref{fig: acoustic field in the glycerol droplet}. As stated in section \ref{sec: Acoust field in the droplet}, it is focused on small regions of the droplets corresponding to the caustics, represented on figure \ref{fig: caustics geometrical interpretation}. As we compare the resulting flow pattern in figure \ref{fig: droplet flow versus visc} to the forcing term, we notice that these caustics act as momentum source-points to generate the flow. For instance, in the glycerol droplet, the forcing terms act on only one side of the droplet and push the flow towards the rear of the droplet. In the case of water, the momentum source terms are more symmetrical and push the fluid in the two opposite directions. Each individual momentum source term results in two eddies, forming the four-swirls pattern. This is particularly visible in figure \ref{fig: PIV bottom view}.

Interestingly, these caustics can be easily constructed from geometrical acoustics. This means that the flow can be, at least qualitatively, predicted from simple geometrical arguments. This assertion must be mitigated by the important role played by the viscosity and the attenuation of sound in the system. 

\subsection{The four-to-two eddies transition}
Since the flow patterns strongly rely on the caustics formation, a key parameter influencing the flow pattern in the drop is the ratio between the droplet diameter and the acoustic wave attenuation length in the fluid:
\begin{equation}
\Lambda = \frac{D}{L_a} = \frac{D \omega^2 \nu b}{c_0^3}
\label{eq: attenuation to droplet diameter}
\end{equation}
For water sessile droplets with a 3.7 mm diameter, $\Lambda \simeq 0.07$ whereas for $90\%$ glycerol droplets $\Lambda \simeq 2.80$. According to the numerical and experimental results, the progressive transition from two to four eddies is located within $0.29 < \Lambda < 1.0 $. In this regard, we understand why a transition of flow pattern happens for this range of viscosity. In glycerol droplets, the sound wave experiences little reflections before fading whereas in water it should bounce at least sixteen times at the droplet interface. Hence, three regimes naturally emerge: $\Lambda << 1$, $\Lambda \simeq 1$ and $\Lambda >> 1$. In the first one, four eddies are formed as it is observed with water. The intermediate regime happens for large glycerol concentrations and results in two vortices. Decreasing further $\Lambda$ was achieved previously by \cite{Beyssen2006} who worked at $40$ MHz frequency with 90 wt$\%$ glycerol mixtures. The resulting flow pattern turns into a single vortex with a horizontal vorticity axis. 

\section{Dimensional analysis}
The previous analysis can be transposed to other frequencies and viscosities by using the Buckingham $\pi$-theorem \cite{bertrand1878,rayleigh1915}. At moderate actuation power (low hydrodynamic Reynolds number), equation (\ref{eqf:streaming}) yields the following scaling for the velocity:
\begin{equation}
\langle V \rangle = f_1\frac{\mathcal{F} D^2}{\mu} .
\label{eq: ND hydrodynamics}
\end{equation}
with $\theta_c$ the contact angle, $\mathcal{F} $ the magnitude of the force, $D$ the droplet diameter and $f_1$ a function of dimensionless parameters. At low Reynolds number, the prefactor $f_1$ depends solely on the droplet geometry, which here can be mainly quantified by the contact angle $\theta_c$ only: $f_1 = f_1(\theta_c)$. The force magnitude $\mathcal{F}$ depends on the acoustic field with input parameters $\tilde{p}$ and $\tilde{v}$, whose magnitude are proportional to $u_0$ (the magnitude of the acoustic perturbation displacement) by linearity and whose topology is determined by the shape of the droplet ($\theta_c$), the dimensionless wavenumber $kD$, the wave radiation angle $\theta_R$ from the substrate to the liquid given by Snell-Descartes law  $\sin(\theta_R)=c_l/c_s$, the characteristic parameter 
of the wave attenuation  in the bulk $\Lambda$, and the characteristic parameter for the surface wave attenuation due to its absorption by the liquid $D\alpha$, with $\alpha$ the attenuation rate introduced in paragraph \ref{sss:msft}. In practice, $\theta_c$ is often chosen near 90$^o$ and $k D << 1$ to optimize streaming efficiency. Consequently, the wave propagation mainly depends on $\Lambda$, $\theta_R$ and $D\alpha$. Most liquid sound speed ranges between 1200 m/s (organic compounds) and 1500 m/s, and SAW are mostly generated on lithium niobate with phase velocity close to 3650 m/s (median velocity of SAW propagating on a lithium niobate X cut over all the directions). This narrows considerably the range of possible Rayleigh angles ($19^o < \theta_R < 24^o$). Hence, the sound propagation in sessile droplets on lithium niobate chiefly depends on $\alpha D$ and $\Lambda$. The force magnitude is then given by:
\begin{equation}
\mathcal{F} = f_2(\Lambda,\alpha D) \frac{\rho_0 \omega^2 {u_0}^2 \Lambda}{D} = f_2(\Lambda,\alpha D) \frac{\rho_0 \omega^4 \nu b  {u_0}^2}{c^3},
\label{eq: ND acoustics}
\end{equation}
where $f_2(\Lambda,\alpha D)$ accounts for the geometrical distribution of the acoustic field. Combining equations (\ref{eq: ND hydrodynamics}) and (\ref{eq: ND acoustics}), and neglecting the influence of the contact angle for the practical reasons detailed previously, we get:
\begin{eqnarray}
\langle V \rangle & = & V_0 f(\Lambda,\alpha D), \\
\label{eq: ND final}
\mbox{ with } V_0 & = & \frac{\omega^4 {u_0}^2 b D^2}{c^3} \mbox{ and  }f(\Lambda,\alpha D) = f_1(\theta_c = 90^o)f_2(\Lambda,\alpha D)
\label{eq: ND magnitude}
\end{eqnarray}
We notice that although the viscosity has no explicit influence on the value of $V_0$, it appears in the expression $\Lambda$ which represents the bulk attenuation of the acoustic wave in the droplet. 

\begin{figure} \begin{center}
\includegraphics[width = 100mm]{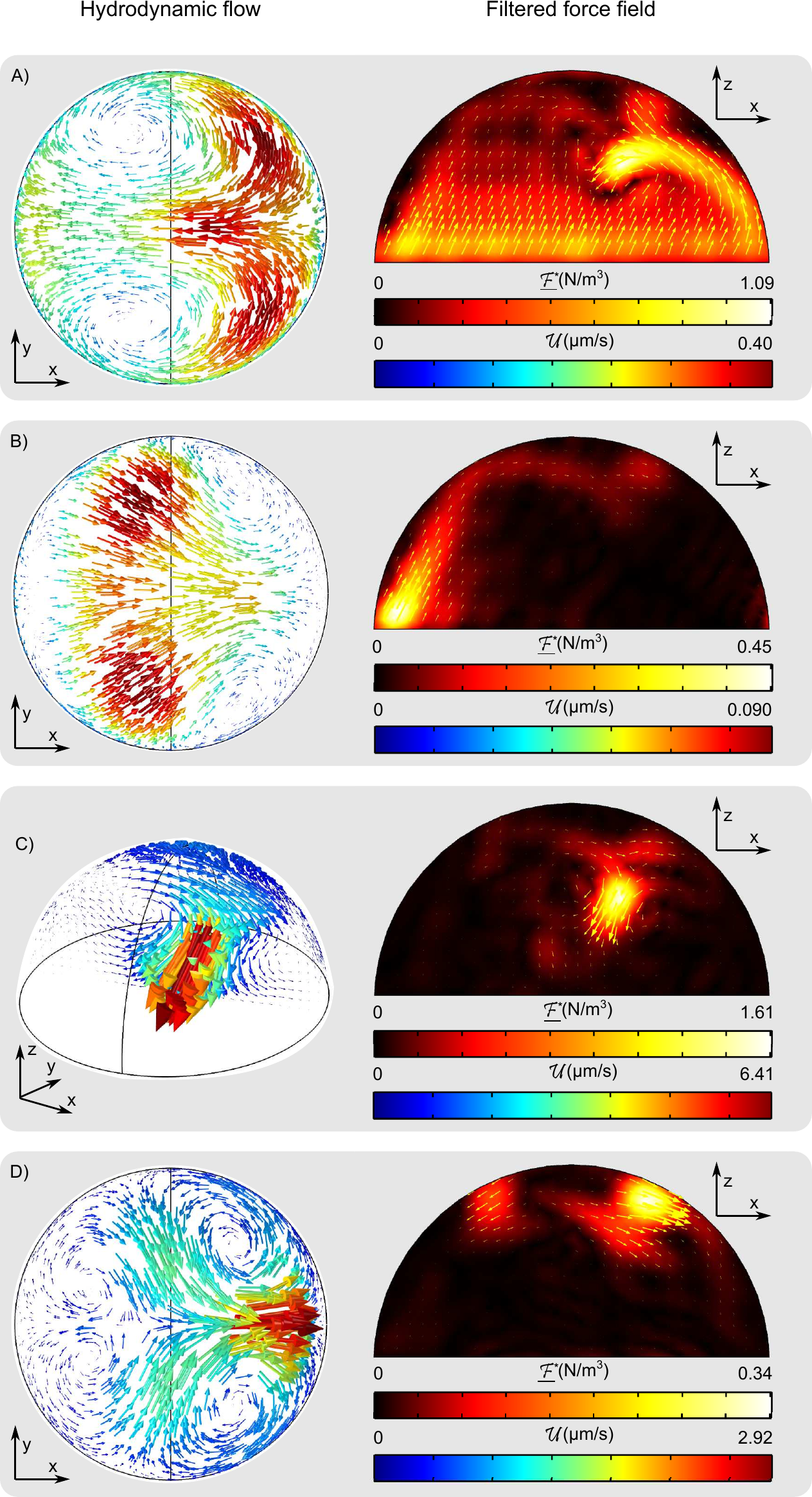}
\caption{Numerical simulation of the four main types of acoustic streaming generated by a plane SAW propagating from left to right. A) $\alpha D=0.17 $, $\Lambda=5.0$ (\textcolor{red}{$\vartriangle$}),  B) $\alpha D=6.0 $, $\Lambda=5.0$ (\textcolor{blue}{O}), C) $\alpha D=0.17 $, $\Lambda=0.27$ (\textcolor{cyan}{$\triangledown$}),  D) $\alpha D=6.0$, $\Lambda=0.27$ (\textcolor{ForestGreen}{$\square$}). Velocity field is depicted on the left, and the forcing term is on the right. Numerical values are computed for an incident SAW of frequency $f=20$MHz and amplitude 10 pm. The droplet diameter is $3.7$ mm, fluid properties are those of water, except for the dynamic viscosity $\mu$ which was used to set $\Lambda$. The other parameter $\alpha D$ was set by tuning the SAW attenuation coefficient in the solid.}
\label{fig: different_streaming}
\end{center} \end{figure}

In order to get a broader picture of the streaming induced by a progressive surface acoustic wave in a sessile droplet, we performed 100 simulations with $\alpha D$ and $\Lambda$ ranging from 0.1 to 10, spanning two orders of magnitude. Depending on the value of these parameters, we observed four distinct streaming flow regimes (see the left column of figure \ref{fig: different_streaming}). At low SAW attenuation and high Bulk Acoustic Wave (BAW) dissipation (small droplet, viscous liquid - A), the flow is driven by two eddies at the front of the droplet. Keeping constant the BAW dissipation but increasing the SAW attenuation (large droplet, viscous liquid - B), the eddies migrate to the droplet rear. At low SAW attenuation and bulk dissipation (small droplet, low viscosity liquid - C), the flow pattern becomes toroidal. Finally, for strong SAW dissipation (large droplet, low viscosity liquid - D), this yields a four-eddies flow field. We related the transition to a switch of relative importance between the incident wave, the primary and the secondary focus (see the right column of figure \ref{fig: different_streaming}). Two topological caustics were deduced from ray acoustics in section \ref{sec: Acoust field in the droplet}. These caustics are fueled by the radiation of surface acoustic waves in the bulk. For a weak SAW attenuation and a strong BAW absorption (A), the force is mainly exerted by the primary focus  (caustic) of the acoustic field, but the surface waves radiating from the front of the droplet are attenuated before reaching the caustic, resulting in a force imbalance that creates the two eddies. Increasing the SAW attenuation, the caustics are less prevalent and the flow is solely driven by the incident SAW (B) which is stronger close to the droplet edge. Decreasing the BAW absorption, the secondary focus (due to reflections guided along the drop free surface) overruns the primary one. Indeed, this focus is fed by the SAW radiating close to the droplet edge and is therefore quite insensitive to the SAW attenuation. At very low BAW absorption like in water at the megahertz range, the waves guided along the droplet surface may bounce multiple times from the left to the right of the droplet, a phenomenon reminiscent of whispering gallery mode guided by the droplet free surface. This results in the symmetrization of the secondary focus and yields a quadripolar flow (D). Finally, decreasing the SAW attenuation, the primary focus dominates again, forming a toroidal flow (C).

\begin{figure} \begin{center}
\includegraphics{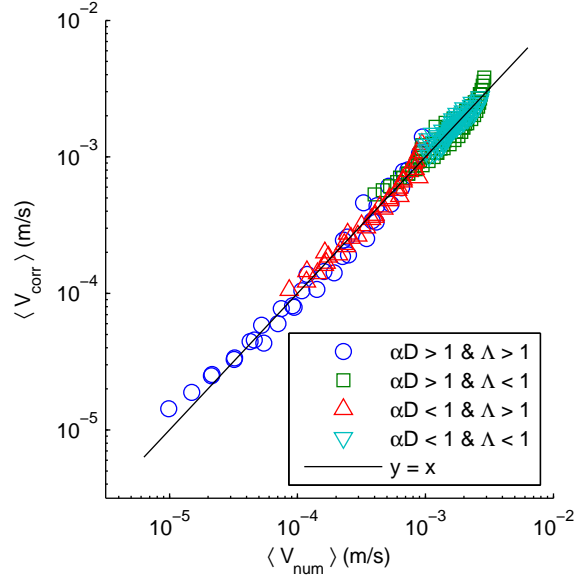}
\caption{Comparison of the average velocity of the droplet internal flow computed from numerical simulations (SSSF) and from the power law given in equation (\ref{eq: ND final}) with the coefficients in table \ref{tab: regression coefficients}.}
\label{fig: check_correlation}
\end{center} \end{figure}

Under the guidance of the numerical results, we looked for a simple expression of the average velocity of the droplet inner flow. This amounts to finding the value of $f$ depending on the two parameters $\alpha D, \Lambda$. We assumed $f$ takes the form of a power law: $f(\alpha D, \Lambda) = k(\alpha D)^a \Lambda^b$ with a set of coefficients $(a,b)$ specific to each of the four flow regimes identified previously. The value of the coefficients $a$ and $b$ indicates the relative importance of the SAW attenuation and the BAW absorption respectively. These coefficients were regressed using multilinear curve fitting, with the resulting values given in table \ref{tab: regression coefficients}. These regression coefficients (table \ref{tab: regression coefficients}) yield fairly accurate average flow velocity as shown in figure \ref{fig: check_correlation}. Interestingly, the average velocity of the flow patterns depicted in figure \ref{fig: different_streaming} C and D shows little dependence on the magnitude of $\Lambda$ ($b\simeq 0.1$) whereas the average flow velocity at higher BAW attenuation is adversely affected by $\Lambda$ and thus by the viscosity. This is in good agreement with Eckart's and Nyborg's view on acoustic streaming. In the former, the wave is assumed to be weakly attenuated over the reservoir extent so the flow velocity is independent of the viscosity. In the latter, the BAW is strongly attenuated within the reservoir length, so the wave momentum is integrally transfered to the fluid. This bounded amount of momentum is in turn dissipated by viscous shear such that increasing viscosity yields lower average velocity with a nearly-linear relationship for small droplets.


\begin{table}
\begin{center}
\begin{tabular}{lccc}
Region  &  $k\times 10^{3}$ & $a$ & $b$ \\
\textcolor{blue}{O}($\alpha D > 1$, $\Lambda > 1$) & 2.06 & -0.54 & -1.60\\
\textcolor{ForestGreen}{$\square$}($\alpha D > 1$, $\Lambda < 1$)  & 1.83  & -0.56 & -0.19 \\
\textcolor{red}{$\vartriangle$}($\alpha D < 1$, $\Lambda > 1$) & 0.75  &   -0.27 &   -0.87 \\
\textcolor{cyan}{$\triangledown$}($\alpha D < 1$, $\Lambda < 1$) & 1.50 & +0.18  & -0.14 \\
\end{tabular}
\end{center}
\caption{Regression coefficients for the average streaming velocity in equation (\ref{eq: ND magnitude}): $f(\Lambda,\alpha D) = k(\alpha D)^a\Lambda^b$ \label{tab: regression coefficients}}

\end{table}

\begin{table}
\begin{center}
\begin{tabular}{cccccc}
$w_\mathtt{glyc.}$ & $\Lambda$  &$\alpha D$ & $ V_0$ (mm/s)& $\langle V_{\mathtt{exp}} \rangle$ &  $\langle V_{\mathtt{corr}} \rangle$  \\
&$\frac{D\omega^2\nu b}{c^3}$& $3.7\frac{\omega \rho_0 D}{10^9}$ & $\frac{\omega^4 u_0^2 b D^2}{c^3}$ & ($\mu$m/s)& ($\mu$m/s)\\
0.00 & 0.068 & 1.7& 17 &  49  & 39 \\
0.10 & 0.080 & 1.8& 16 &  43  & 35 \\
0.20 & 0.095 & 1.8& 15 &  24  & 31 \\
0.30 & 0.11  & 1.9& 13 &  25  & 26 \\
0.40 & 0.15  & 1.9& 12 &  20  & 22 \\ 
0.50 & 0.20  & 2.0& 10 &  11  & 17\\
0.60 & 0.29  & 2.1& 9.0 & 7.4 & 14 \\
0.70 & 0.52  & 2.1& 8.0 & 2.7 & 11\\
0.80 & 1.0   & 2.2& 6.8 & 2.6 & 8.4\\
0.90 & 2.80  & 2.3& 5.4 & 0.94& 1.4 \\
\end{tabular}
\end{center}
\caption{Non-dimensional parameters extracted from table \ref{tab: physical parameters}. $\langle V \rangle_{\mathtt{exp}}$ is the average velocity in the droplet measured from the experiments described in section \ref{sec: Experiment}, and $\langle V \rangle_{\mathtt{corr}}$ is computed from equation (\ref{eq: ND final}) with the coefficients given in table \ref{tab: regression coefficients}.  \label{tab: adimensional}}
\end{table}

We also provide a comparison with experimental data in table \ref{tab: adimensional}. These correlations are not limited to 20 MHz. For instance, in the 6 MHz simulation of section \ref{sec: DNS_vs_SSSF}, $\alpha D = 0.27$, $\Lambda = 0.004$ and $V_0 = 1.0$ m/s, yielding $\langle V \rangle = 2.6$ mm/s compared to 2.86 and 1.92 mm/s obtained numerically depending on the model. This indicates that full similitude requirements (especially the non-dimensional acoustic wavelength $k D$) are not mandatory to obtain quantitative results of acoustic streaming. Instead, our analysis provides some guidelines for the study of more complex acoustic fields: the most relevant parameters for a partial similitude computation of sessile droplet streaming are the BAW and SAW attenuation $\Lambda$ and $\alpha D$. Consequently, high frequency acoustic SAW induced streaming can be conveniently simulated at a few megahertz with a direct numerical simulation method, and then extrapolated to other cases with identical $\alpha D$ and $\Lambda$ while at higher frequencies or droplet sizes.

\section{Conclusion}
In this paper, we investigated thoroughly the phenomenon of bulk (Eckart) acoustic streaming in hemispherical cavities much larger than the acoustic wavelength. This study is especially relevant for the MHz actuation of sessile drops with SAW, which served as cased study and experimental check. 

The main issues addressed by this study are the measurement of the acoustic and hydrodynamic fields in the drop and the simulation of systems with large characteristic size compared to the wavelength, which may lead to prohibitive computation time. In this paper, we presented a numerical method that reduces considerably the computational costs and thus enables such complex simulations on a desktop computer. From a physical perspective, we have shown that contrarily to a widespread belief, the viscosity plays a major role on the acoustic streaming in cavities. This effect is demonstrated both numerically and experimentally in sessile droplets excited by surface acoustic waves. The experiments were also used to validate our numerical scheme, which in turn became a valuable tool to visualize the acoustic field in the drop and unveil the spatial distribution of the forcing term. It turns out that the streaming force is mainly concentrated in some caustics whose position can be obtained easily from geometrical acoustics. This knowledge supports the possibility to predict qualitatively the flow pattern in large objects from geometrical acoustics.

A possible continuation of this work would then be to implement a fast ray-acoustic model and use the tremendous progress recently accomplished in the field of caustics to get explicitly the force terms acting on the fluid. This would preserve much time to perform ambitious simulations with moving boundaries as observed in experiments involving high acoustic power, including droplet deformation, displacement, mixing, heating and atomization. Aside from the fundamental questions related to droplet acoustofluidics answered by the present study, we reduced the computation of acoustic streaming in sessile droplets to two non-dimensional parameters. This allows to study high frequency SAW droplet actuation (not restricted to plane waves) based on a partial similitude approach, simply by extrapolating the quantitative flow pattern obtained at lower SAW frequency simulations. Such an approach allows using DNS strategies for convenient code development while keeping reasonable memory requirements.

\begin{acknowledgments}
We gratefully acknowledge Tiesse Diarra for his important contribution in writing the azimuthal Fourier transform algorithm essential to this simulation. We also thanks Bernard Bonello for his support in providing the workstation to perform the simulations. This work is supported by
ANR Project No. ANR-12-BS09-0021-01, ANR-12-BS09-0021-02, and R\'{e}gion Nord Pas de Calais.
\end{acknowledgments}

\appendix

\section{Computational cost} \label{sec: comput_cost}
\begin{figure} \begin{center}
\includegraphics{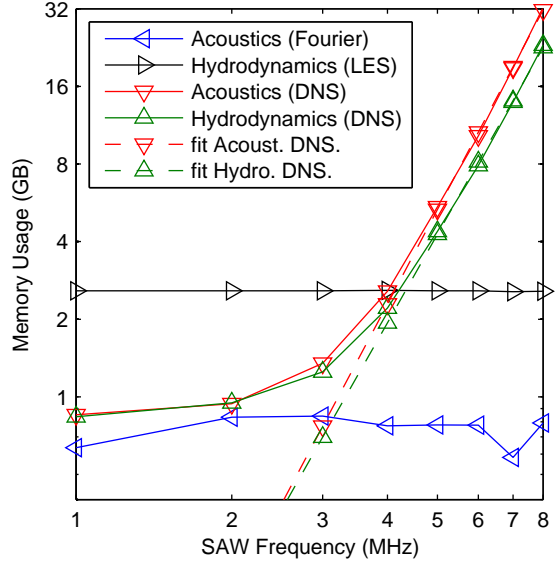}
\caption{Memory usage versus SAW frequency for the various parts and methods of the computation. Regression coefficients are $M_\mathtt{Acoust DNS} = 0.012\times F^{3.79}$ and $M_\mathtt{Hydro DNS} = 0.014\times F^{3.54}$ where F is the SAW frequency in MHz and M the memory in GB. Projections at 20 MHz indicate 1.0 TB of RAM of acoustics (DNS) and 580 GB of RAM for the fluidics (DNS).}
\label{fig: memory DNS SSSF}
\end{center} \end{figure}

Computing the acoustic streaming might seem to be an easy task with appropriate softwares. Indeed, many codes are readily available to compute acoustic fields and fluid mechanics. The nonlinear hydrodynamic forcing term can be deduced from the acoustics and computed in a straightforward fashion. Nevertheless, in high frequency regimes (with wavelength much smaller than the characteristic length scale of the flow structured produced), the computation time can become prohibitive. In figure \ref{fig: memory DNS SSSF}, we compare the memory requirements of the direct numerical simulations as already implemented in the commercial software Comsol 4.3b to our more customized implementations. The benchmark test is a 2 mm diameter sessile droplet (water, contact angle 100$^o$) exposed to an incident SAW radiation for a range of megahertz frequencies.

We first notice that there is some background noise on the memory requirements, which magnitude is about 700 MB, probably related to the OS (Windows 7) and the software. The direct numerical simulations of acoustics and hydrodynamics start consuming a lot of memory after a 4 MHz threshold. This corresponds to a wavelength of 375 $\mu$m, which is a third of the droplet radius. After this threshold, the memory requirements grow quickly and extrapolation to 20 MHz excitation estimate the need to 1 TB for the acoustics, and 580 GB for the fluidics. Access to such middle range cluster capabilities being difficult, we used alternative numerical recipes.

The Fourier transform resolves the incident field into azimuthal harmonics to reduce the computation of acoustics to a 2D problem. The memory requirements at these excitation frequency are so low that they are overwhelmed by the background noise. 

The Large Eddy Simulation method with filtered source term is always computed with the same grid resolution, which is fixed by the explicit filtering step. The memory needed to compute the force is not shown since it is implementation-dependent. With Matlab, we reconstructed the 3D acoustic variables in multidimensional arrays $\tilde{p}$, $\tilde{v}_r$, $\tilde{v}_\theta$ and $\tilde{v}_z$, each of which weighs about 190 MB for computations of 20 MHz acoustic fields. We used the multidimensional Fourier transform in Matlab to maximize the speed when filtering the forces and memory requirements always kept below 32 GB even at 20 MHz. If required, the filtering can be achieved in the real space with low memory consumption by using cross correlation algorithm to smooth directly the force field. 

\bibliographystyle{jfm}

\bibliography{ViscosityStreaming,biblioa}

\begin{thebibliography}{66}
\expandafter\ifx\csname natexlab\endcsname\relax\def\natexlab#1{#1}\fi

\bibitem[Alghane {\em et~al.\/}(2012)Alghane, Chen, Fu, Li, Desmulliez,
  Mohammed \& Walton]{Alghane2012}
{\sc Alghane, M., Chen, B.~X., Fu, Y.~Q., Li, Y., Desmulliez, M. P.~Y.,
  Mohammed, M.~I. \& Walton, A.~J.} 2012 Nonlinear hydrodynamic effects induced
  by rayleigh surface acoustic wave in sessile droplets. {\em Phys. Rev. E\/}
  {\bf 86}, 056304.

\bibitem[Alghane {\em et~al.\/}(2011)Alghane, Chen, Fu, Li, Luo \&
  Walton]{Alghane2011}
{\sc Alghane, M, Chen, B~X, Fu, Y~Q, Li, Y, Luo, J~K \& Walton, A~J} 2011
  Experimental and numerical investigation of acoustic streaming excited by
  using a surface acoustic wave device on a 128$^o$ yx-linbo 3 substrate. {\em
  J. Micromech. Microeng.\/} {\bf 21}~(1), 015005.

\bibitem[Alzuaga {\em et~al.\/}(2005)Alzuaga, Manceau \& Bastien]{Alzuaga2005}
{\sc Alzuaga, S., Manceau, J.-F. \& Bastien, F.} 2005 Motion of droplets on
  solid surface using acoustic radiation pressure. {\em J. Sound. Vib.\/} {\bf
  282}~(1–2), 151 -- 162.

\bibitem[Baudoin {\em et~al.\/}(2012)Baudoin, Brunet, Bou~Matar \&
  Herth]{apl_baudoin_2012}
{\sc Baudoin, M., Brunet, P., Bou~Matar, O. \& Herth, E.} 2012 Low energy
  droplet actuation via modulated surface acoustic waves. {\em Appl. Phys.
  Lett.\/} {\bf 100}, 154102.

\bibitem[Bertrand(1878)]{bertrand1878}
{\sc Bertrand, J} 1878 Sur l'homog{\'e}n{\'e}it{\'e} dans les formules de
  physique. {\em Comptes rendus\/} {\bf 86}~(15), 916--920.

\bibitem[Beyssen {\em et~al.\/}(2006)Beyssen, Le~Brizoual, Elmazria, Alnot,
  Perry \& Maillet]{Beyssen2006}
{\sc Beyssen, D., Le~Brizoual, L., Elmazria, O., Alnot, P., Perry, I. \&
  Maillet, D.} 2006 6i-2 droplet heating system based on saw/liquid
  interaction. In {\em IEEE Ultras. Symp.\/}, pp. 949--952.

\bibitem[Bou-Zeid(2015)]{Bou-Zeid2015}
{\sc Bou-Zeid, Elie} 2015 Challenging the large eddy simulation technique with
  advanced a posteriori tests. {\em J. Fluid Mech.\/} {\bf 764}, 1--4.

\bibitem[Bourquin {\em et~al.\/}(2010)Bourquin, Reboud, Wilson \&
  Cooper]{Bourquin2010}
{\sc Bourquin, Yannyk, Reboud, Julien, Wilson, Rab \& Cooper, Jonathan~M.} 2010
  Tuneable surface acoustic waves for fluid and particle manipulations on
  disposable chips. {\em Lab Chip\/} {\bf 10}, 1898--1901.

\bibitem[Brunet {\em et~al.\/}(2010)Brunet, Baudoin, Bou~Matar \&
  Zoueshtiagh]{pre_brunet_2010}
{\sc Brunet, P., Baudoin, M., Bou~Matar, O. \& Zoueshtiagh, F.} 2010 Droplet
  displacement and oscillations induced by ultrasonic surface acoustic waves: a
  quantitative study. {\em Phys. Rev. E\/} {\bf 81}, 036315.

\bibitem[B{\"u}hler(2009)]{Buhler2009}
{\sc B{\"u}hler, O.} 2009 {\em Waves and mean flows\/}. Cambridge University
  Press.

\bibitem[Campbell \& Jones(1970)]{CampbellJones}
{\sc Campbell, J.~J. \& Jones, W.~R.} 1970 Propagation of surface waves at the
  boundary between a piezoelectric crystal and a fluid medium. {\em IEEE T.
  Ultrason. Ferr.\/} {\bf 17}~(2), 71--76.

\bibitem[Cheeke(2002)]{Cheeke}
{\sc Cheeke, J. D.~N.} 2002 {\em Fundamental and Applications of Ultrasonic
  Waves\/}, p. 8.3. CRC Press LLC.

\bibitem[Cheng(2008)]{Cheng2008}
{\sc Cheng, Nian-Sheng} 2008 Formula for the viscosity of a glycerol-water
  mixture. {\em Ind. Eng. Chem. Res.\/} {\bf 47}~(9), 3285--3288.

\bibitem[Collignon {\em et~al.\/}(2015)Collignon, Friend \& Yeo]{Collignon2015}
{\sc Collignon, Sean, Friend, James \& Yeo, Leslie} 2015 Planar microfluidic
  drop splitting and merging. {\em Lab Chip\/} {\bf 15}, 1942--1951.

\bibitem[Coulouvrat(1992)]{ja_coulouvrat_1992}
{\sc Coulouvrat, F.} 1992 On the equations of nonlinear acoustics. {\em J.
  acoust.\/} {\bf 5}, 321--359.

\bibitem[Deardorff(1970)]{Deardorff1970}
{\sc Deardorff, James~W.} 1970 A numerical study of three-dimensional turbulent
  channel flow at large reynolds numbers. {\em J. Fluid Mech.\/} {\bf 41},
  453--480.

\bibitem[Dentry {\em et~al.\/}(2014)Dentry, Yeo \& Friend]{pre_dentry_2014}
{\sc Dentry, M.~B., Yeo, L.Y. \& Friend, J.R.} 2014 Frequency effects on the
  scale and behavior of acoustic streaming. {\em Phys. Rev. E\/} {\bf 89},
  013203.

\bibitem[Du {\em et~al.\/}(2009)Du, Swanwick, Fu, Luo, Flewitt, Lee, Maeng \&
  Milne]{Du2009}
{\sc Du, X~Y, Swanwick, M~E, Fu, Y~Q, Luo, J~K, Flewitt, A~J, Lee, D~S, Maeng,
  S \& Milne, W~I} 2009 Surface acoustic wave induced streaming and pumping in
  128$^o$ y-cut linbo 3 for microfluidic applications. {\em J. Micromech.
  Microeng.\/} {\bf 19}~(3), 035016.

\bibitem[Eckart(1948)]{Eckart}
{\sc Eckart, C.} 1948 Vortices and streams caused by sound waves. {\em Phys.
  Rev.\/} {\bf 73}, 68--76.

\bibitem[Friend \& Yeo(2011)]{rmp_friend_2011}
{\sc Friend, J.R. \& Yeo, L.Y.} 2011 Microscale acoustofluidics: Microfluidics
  driven via acoustics and ultrasonics. {\em Rev. Mod. Phys.\/} {\bf 83},
  647--704.

\bibitem[Frommelt {\em et~al.\/}(2008)Frommelt, Kostur, Talkner, H\"{a}nggi \&
  Wixforth]{prl_frommelt_2008}
{\sc Frommelt, T., Kostur, M.and Wenzel-Sch\"{a}fer, M., Talkner, P.,
  H\"{a}nggi, P. \& Wixforth, A.} 2008 Microfluidic mixing via acoustically
  driven chaotic advection. {\em Phys. Rev. Lett.\/} {\bf 100}, 034502.

\bibitem[Fukaya \& Kondoh(2015)]{Tomohiko2015}
{\sc Fukaya, Tomohiko \& Kondoh, Jun} 2015 Experimental consideration of
  droplet manipulation mechanism using surface acoustic wave. {\em Jpn. J.
  Appl. Phys.\/} {\bf 54}~(7S1), 07HE06.

\bibitem[Gusev \& Rudenko(1979)]{Gusev1979}
{\sc Gusev, V.~E. \& Rudenko, O.~V.} 1979 Nonstready quasi-one-dimensional
  acoustic streaming in unbounded volumes with hydrodynamic nonlinearity. {\em
  Sov. Phys. Acoust.\/} {\bf 25}, 493--497.

\bibitem[Hertz \& Mende(1939)]{Hertz_Mende}
{\sc Hertz, G. \& Mende, H.} 1939 Der strahlungsdrunk in fl{\"u}ssigkeiten.
  {\em Z. Phys.\/} {\bf 114}, 354–367.

\bibitem[Ito {\em et~al.\/}(2007)Ito, Sugimoto, Matsui \&
  Kondoh]{jjap_ito_2007}
{\sc Ito, S., Sugimoto, Y., Matsui, Y. \& Kondoh, J.} 2007 Study of surface
  acoustic wave streaming phenomenon based on temperature measurement and
  observation of streaming in liquids. {\em Jpn. J. Appl. Phys.\/} {\bf 46},
  4718.

\bibitem[Kamakura {\em et~al.\/}(1995)Kamakura, Matsuda, Kumamoto \&
  Breazeale]{Kamakura1995}
{\sc Kamakura, Tomoo, Matsuda, Kazuhisa, Kumamoto, Yoshiro \& Breazeale,
  Mack~A.} 1995 Acoustic streaming induced in focused gaussian beams. {\em J.
  Acoust. Soc. Am.\/} {\bf 97}~(5), 2740--2746.

\bibitem[Kondoh {\em et~al.\/}(2005)Kondoh, Shimizu, Matsui, Sugimoto \&
  Shiokawa]{ieeeus_kondoh_2005}
{\sc Kondoh, J., Shimizu, N., Matsui, Y., Sugimoto, M. \& Shiokawa, S.} 2005
  Development of saw thermocycler for small liquid droplets. {\em IEEE
  Ultrason. Symp.\/} {\bf 2}, 1023--1027.

\bibitem[Kondoh {\em et~al.\/}(2009)Kondoh, Shimizu, Matsui, Sugimoto \&
  Shiokawa]{saa_kondoh_2009}
{\sc Kondoh, J, Shimizu, N., Matsui, Y., Sugimoto, M. \& Shiokawa, S.} 2009
  Development of temperature control system for liquid droplet using surface
  acoustic wave device. {\em Sensors and actuators A\/} {\bf 149}, 292--297.

\bibitem[K{\"o}ster(2007)]{koster2007}
{\sc K{\"o}ster, Daniel} 2007 Numerical simulation of acoustic streaming on
  surface acoustic wave-driven biochips. {\em SIAM Journal on Scientific
  Computing\/} {\bf 29}~(6), 2352--2380.

\bibitem[Kuznetsov(1970)]{spa_kuznetsov_1970}
{\sc Kuznetsov, V.P.} 1970 Equations of nonlinear acoustics. {\em Sov. Phys.
  Acoust.\/} {\bf 16}, 467--470.

\bibitem[Liebermann(1949)]{Liebermann1949}
{\sc Liebermann, L.~N.} 1949 The second viscosity of liquids. {\em Phys.
  Rev.\/} {\bf 75}, 1415--1422.

\bibitem[Lighthill(1978)]{Lighthill1978391}
{\sc Lighthill, Sir~James} 1978 Acoustic streaming. {\em J. Sound Vibr.\/} {\bf
  61}~(3), 391 -- 418.

\bibitem[{Lord Rayleigh}(1884)]{ptrsl_rayleigh_1884}
{\sc {Lord Rayleigh}} 1884 On the circuclation of air observed in {K}undt's
  tubes, ans some allied acoustical problems. {\em Ph. Trans. Roy Soc.
  London\/} {\bf 175}, 1--21.

\bibitem[Matsuda {\em et~al.\/}(1996)Matsuda, Kamakura \&
  Kumamoto]{Matsuda1996763}
{\sc Matsuda, Kazuhisa, Kamakura, Tomoo \& Kumamoto, Yoshiro} 1996 Buildup of
  acoustic streaming in focused beams. {\em Ultrasonics\/} {\bf 34}~(7), 763 --
  765.

\bibitem[Mitome(1998)]{Mitome1998}
{\sc Mitome, Hideto} 1998 The mechanism of generation of acoustic streaming.
  {\em Electronics and Communications in Japan (Part III: Fundamental
  Electronic Science)\/} {\bf 81}~(10), 1--8.

\bibitem[Nyborg(1953)]{ap_nyborg_1998}
{\sc Nyborg, Wesley~L.} 1953 Acoustic streaming due to attenuated plane waves.
  {\em The Journal of the Acoustical Society of America\/} {\bf 25}~(1),
  68--75.

\bibitem[Pope(2004)]{Pope2004}
{\sc Pope, Stephen~B} 2004 Ten questions concerning the large-eddy simulation
  of turbulent flows. {\em New J. Phys.\/} {\bf 6}~(1), 35.

\bibitem[Qi {\em et~al.\/}(2008)Qi, Yeo \& Friend]{pof_yeo_2008}
{\sc Qi, A., Yeo, L.Y. \& Friend, J.R.} 2008 Interfacial destabilization and
  atomization driven by surface acoustic waves. {\em Phys. Fluids\/} {\bf
  20}~(7), 074103.

\bibitem[Quintero \& Simonetti(2013)]{Quintero2013}
{\sc Quintero, R. \& Simonetti, F.} 2013 Rayleigh wave scattering from sessile
  droplets. {\em Phys. Rev. E\/} {\bf 88}, 043011.

\bibitem[Raghavan {\em et~al.\/}(2010)Raghavan, Friend \& Yeo]{Raghavan2010}
{\sc Raghavan, Rohan~V., Friend, James~R. \& Yeo, Leslie~Y.} 2010 Particle
  concentration via acoustically driven microcentrifugation: micropiv flow
  visualization and numerical modelling studies. {\em Microfluid. Nanofluid.\/}
  {\bf 8}~(1), 73--84.

\bibitem[Rayleigh(1915)]{rayleigh1915}
{\sc Rayleigh, Lord} 1915 The principle of similitude. {\em Nature\/} {\bf
  95}~(66), 591.

\bibitem[Reboud {\em et~al.\/}(2012)Reboud, Bourquin, Wilson, Pall, Jiwaji,
  Pitt, Graham, Waters \& Cooper]{pnas_reboud_2012}
{\sc Reboud, J., Bourquin, Y., Wilson, G.S., Pall, G.S., Jiwaji, M., Pitt,
  A.R., Graham, A., Waters, A.P. \& Cooper, J.M.} 2012 Shaping acoustic fields
  as a toolset for microfluidic manipulations in diagnostic technologies. {\em
  Proc. Nat. Ac. Sci. USA\/} {\bf 109}, 15162.

\bibitem[Rednikov \& Sadhal(2011)]{Rednikov2011}
{\sc Rednikov, A.~Y. \& Sadhal, S.~S.} 2011 Acoustic/steady streaming from a
  motionless boundary and related phenomena: generalized treatment of the inner
  streaming and examples. {\em J. Fluid Mech.\/} {\bf 667}, 426--462.

\bibitem[Renaudin {\em et~al.\/}(2006)Renaudin, Tabourier, Zang, Camart \&
  Druon]{saa_renaudin_2006}
{\sc Renaudin, A., Tabourier, P., Zang, V., Camart, J.C. \& Druon, C.} 2006 Saw
  nanopump for handling droplets in view of biological applications. {\em
  Sensors and Actuators B\/} {\bf 113}, 389--397.

\bibitem[Riaud {\em et~al.\/}(2015)Riaud, Baudoin, Thomas \&
  Bou~Matar]{ieeetuffc_riaud_2016}
{\sc Riaud, A., Baudoin, M., Thomas, J.L. \& Bou~Matar, O.} 2015 {SAW}
  synthesis with {IDT}s array and the inverse filter: toward a versatile saw
  toolbox for microfluidics and biological applications. {\em submitted to IEEE
  T. Ultrason. Ferr.\/} .

\bibitem[Riley(1998)]{Riley_TCFD}
{\sc Riley, N.} 1998 Acoustic streaming. {\em Theor. Comp. Fluid Dyn.\/} {\bf
  10}~(1-4), 349--356.

\bibitem[Riley(2001)]{arfm_ryley_2001}
{\sc Riley, N.} 2001 Steady streaming. {\em Ann. Rev. Fluid Mech.\/} {\bf 33},
  43--65.

\bibitem[Rooney {\em et~al.\/}(1982)Rooney, Smith \& Carey]{Rooney1982}
{\sc Rooney, J.~A., Smith, C.~W. \& Carey, R.~F.} 1982 Acoustic streaming in
  superfluid helium. {\em J. Acoust. Soc. Am.\/} {\bf 72}~(1), 245--249.

\bibitem[Roux-Marchand {\em et~al.\/}(2015)Roux-Marchand, Beyssen, Sarry \&
  Elmazria]{ieeetuffc_rouxmarchand_2015}
{\sc Roux-Marchand, T., Beyssen, D., Sarry, F. \& Elmazria, O.} 2015 Rayleigh
  surface acoustic waves as an efficient heating system for biological
  reactions: investigation of microdroplet temperature uniformity. {\em IEEE
  Trans. Ultras. Ferroelec. Freq. Contr.\/} {\bf 62}~(4), 729--735.

\bibitem[Roux-Marchand {\em et~al.\/}(2012)Roux-Marchand, Beyssen, Sarry,
  Grandemange \& Elmazria]{Roux-Marchand2012}
{\sc Roux-Marchand, T., Beyssen, D., Sarry, F., Grandemange, S. \& Elmazria,
  O.} 2012 Microfluidic heater assisted by rayleigh surface acoustic wave on
  aln/128$^o$ y-x linbo3 multilayer structure. In {\em IEEE Int. Ultras.
  Symp.\/}, pp. 1706--1709.

\bibitem[Royer \& Dieulesaint(1996)]{Royer1}
{\sc Royer, Daniel \& Dieulesaint, Eugene} 1996 {\em Elastic waves in solids
  1\/}. Springer.

\bibitem[Royer \& Dieulesaint(1999)]{Royer2}
{\sc Royer, Daniel \& Dieulesaint, Eugene} 1999 {\em Elastic waves in solids
  2\/}. Springer.

\bibitem[Sato \& Fujii(2001)]{Sato_and_Fujii}
{\sc Sato, Masanori \& Fujii, Toshitaka} 2001 Quantum mechanical representation
  of acoustic streaming and acoustic radiation pressure. {\em Phys. Rev. E\/}
  {\bf 64}, 026311.

\bibitem[Schindler {\em et~al.\/}(2006)Schindler, Talkner \&
  H{\"a}nggi]{Schindler2006}
{\sc Schindler, Michael, Talkner, Peter \& H{\"a}nggi, Peter} 2006 Computing
  stationary free-surface shapes in microfluidics. {\em Physics of Fluids
  (1994-present)\/} {\bf 18}~(10), 103303.

\bibitem[Shilton {\em et~al.\/}(2015)Shilton, Mattoli, Travagliati, Agostini,
  Desii, Beltram \& Cecchini]{afm_shilton_2015}
{\sc Shilton, R.~J., Mattoli, V., Travagliati, M., Agostini, M., Desii, A.,
  Beltram, F. \& Cecchini, M.} 2015 Rapid and controllable digital microfluidic
  heating by surface acoustic waves. {\em Adv. Funct. Mat.\/} {\bf 25},
  5895--5901.

\bibitem[Shiokawa {\em et~al.\/}(1990)Shiokawa, Matsui \&
  Ueda]{jjap_shiokawa_1990}
{\sc Shiokawa, S., Matsui, Y. \& Ueda, T.} 1990 Study on {SAW} streaming and
  its application to fluid device. {\em Jpn. J. Appl. Phys.\/} {\bf 29}~(Sup.
  29-1), 137--139.

\bibitem[Slie {\em et~al.\/}(1966)Slie, Donfor \& Litovitz]{Slie1966}
{\sc Slie, W.~M., Donfor, A.~R. \& Litovitz, T.~A.} 1966 Ultrasonic shear and
  longitudinal measurements in aqueous glycerol. {\em J. Chem. Phys.\/} {\bf
  44}~(10), 3712--3718.

\bibitem[Sritharan {\em et~al.\/}(2006)Sritharan, Strobl, Schneider \&
  Wixforth]{apl_sritharan_2006}
{\sc Sritharan, K., Strobl, C.J., Schneider, M.F. \& Wixforth, A.} 2006
  Acoustic mixing at low reynolds numbers. {\em Appl. Phys. Lett.\/} p. 054102.

\bibitem[Stanzial {\em et~al.\/}(2003)Stanzial, Bonsi \&
  Schiffrer]{stanzial2003}
{\sc Stanzial, D, Bonsi, D \& Schiffrer, G} 2003 Four-dimensional treatment of
  linear acoustic fields and radiation pressure. {\em Acta Acust. United.
  Ac.\/} {\bf 89}~(2), 213--224.

\bibitem[Tan {\em et~al.\/}(2009)Tan, Friend \& Yeo]{prl_tan_2009}
{\sc Tan, M.K., Friend, J.R. \& Yeo, L.Y.} 2009 Interfacial jetting phenomena
  induced by focused surface vibrations. {\em Phys. Rev. Lett.\/} {\bf
  103}~(2), 024501.

\bibitem[Tanter {\em et~al.\/}(2001)Tanter, Thomas, Coulouvrat \&
  Fink]{pre_tanter_2001}
{\sc Tanter, M., Thomas, J.L., Coulouvrat, F. \& Fink, M.} 2001 Breaking of the
  time reversal invariance in nonlinear acoustics. {\em Phys. Rev. E\/} p.
  016602.

\bibitem[Vanneste \& B{\"u}hler(2011)]{Vanneste2011}
{\sc Vanneste, J. \& B{\"u}hler, O.} 2011 Streaming by leaky surface acoustic
  waves. {\em P. Roy. soc. A-Math Phy.\/} {\bf 467}~(2130), 1779--1800.

\bibitem[Westervelt(1953)]{Westerwelt1953}
{\sc Westervelt, Peter~J.} 1953 The theory of steady rotational flow generated
  by a sound field. {\em J. Acoust. Soc. Am.\/} {\bf 25}~(1), 60--67.

\bibitem[Wiklund(2012)]{loc_wiklund_2012}
{\sc Wiklund, M.} 2012 Acoustofluidics 14: Applications of acoustic streaming
  in microfluidic devices. {\em Lab Chip\/} {\bf 12}, 2438--2451.

\bibitem[Wixforth {\em et~al.\/}(2004)Wixforth, Strobl, Gauer, Toegl, Scriba \&
  Guttenberg]{abc_wixforth_2004}
{\sc Wixforth, A., Strobl, C., Gauer, C., Toegl, A., Scriba, J. \& Guttenberg,
  Z.} 2004 Acoustic manipulation of small droplets. {\em Anal. Bioanal.
  Chem.\/} {\bf 113}.

\bibitem[Zhang {\em et~al.\/}(2013)Zhang, Zha \& Fu]{Zhang_Anliang2013}
{\sc Zhang, Anliang, Zha, Yan \& Fu, Xingting} 2013 Splitting a droplet with
  oil encapsulation using surface acoustic wave excited by electric signal with
  low power. {\em AIP Advances\/} {\bf 3}~(7), 072119.

\end{thebibliography}

\end{document}